\documentclass[12pt]{article}
\usepackage{color}
\usepackage{epsfig,feynmf}

\makeatletter
\@addtoreset{equation}{section}
\renewcommand{\theequation}{\thesection.\arabic{equation}}
\renewcommand{\thefootnote}{\fnsymbol{footnote}}
\makeatother

\begin{document}
\newcommand{\p}[1]{(\ref{#1})}
\newcommand {\bea}{\begin{eqnarray}}
\newcommand {\eea}{\end{eqnarray}}
\newcommand {\non}{\nonumber\\}
\newcommand {\eq}[1]{\label {eq.#1}}
\newcommand {\defeq}{\stackrel{\rm def}{=}}
\newcommand {\gto}{\stackrel{g}{\to}}
\newcommand {\hto}{\stackrel{h}{\to}}
\newcommand {\1}[1]{\frac{1}{#1}}
\newcommand {\2}[1]{\frac{i}{#1}}
\newcommand {\thb}{\bar{\theta}}
\newcommand {\ps}{\psi}
\newcommand {\psb}{\bar{\psi}}
\newcommand {\ph}{\varphi}
\newcommand {\phs}[1]{\varphi^{*#1}}
\newcommand {\sig}{\sigma}
\newcommand {\sigb}{\bar{\sigma}}
\newcommand {\Ph}{\Phi}
\newcommand {\Phd}{\Phi^{\dagger}}
\newcommand {\Sig}{\Sigma}
\newcommand {\Phm}{{\mit\Phi}}
\newcommand {\eps}{\varepsilon}
\newcommand {\del}{\partial}
\newcommand {\dagg}{^{\dagger}}
\newcommand {\pri}{^{\prime}}
\newcommand {\prip}{^{\prime\prime}}
\newcommand {\pripp}{^{\prime\prime\prime}}
\newcommand {\prippp}{^{\prime\prime\prime\prime}}
\newcommand {\pripppp}{^{\prime\prime\prime\prime\prime}}
\newcommand {\delb}{\bar{\partial}}
\newcommand {\zb}{\bar{z}}
\newcommand {\mub}{\bar{\mu}}
\newcommand {\nub}{\bar{\nu}}
\newcommand {\lam}{\lambda}
\newcommand {\lamb}{\bar{\lambda}}
\newcommand {\kap}{\kappa}
\newcommand {\kapb}{\bar{\kappa}}
\newcommand {\xib}{\bar{\xi}}
\newcommand {\ep}{\epsilon}
\newcommand {\epb}{\bar{\epsilon}}
\newcommand {\Ga}{\Gamma}
\newcommand {\rhob}{\bar{\rho}}
\newcommand {\etab}{\bar{\eta}}
\newcommand {\chib}{\bar{\chi}}
\newcommand {\tht}{\tilde{\th}}
\newcommand {\zbasis}[1]{\del/\del z^{#1}}
\newcommand {\zbbasis}[1]{\del/\del \bar{z}^{#1}}
\newcommand {\vecv}{\vec{v}^{\, \prime}}
\newcommand {\vecvd}{\vec{v}^{\, \prime \dagger}}
\newcommand {\vecvs}{\vec{v}^{\, \prime *}}
\newcommand {\alpht}{\tilde{\alpha}}
\newcommand {\xipd}{\xi^{\prime\dagger}}
\newcommand {\pris}{^{\prime *}}
\newcommand {\prid}{^{\prime \dagger}}
\newcommand {\Jto}{\stackrel{J}{\to}}
\newcommand {\vprid}{v^{\prime 2}}
\newcommand {\vpriq}{v^{\prime 4}}
\newcommand {\vt}{\tilde{v}}
\newcommand {\vecvt}{\vec{\tilde{v}}}
\newcommand {\vecpht}{\vec{\tilde{\phi}}}
\newcommand {\pht}{\tilde{\phi}}
\newcommand {\goto}{\stackrel{g_0}{\to}}
\newcommand {\tr}{{\rm tr}\,}
\newcommand {\GC}{G^{\bf C}}
\newcommand {\HC}{H^{\bf C}}
\newcommand{\vs}[1]{\vspace{#1 mm}}
\newcommand{\hs}[1]{\hspace{#1 mm}}
\newcommand{\al}{\alpha}
\newcommand{\be}{\beta}
\newcommand{\Lam}{\Lambda}
\newcommand{\kahler}{K\"ahler }
\newcommand{\con}[1]{{\Gamma^{#1}}}
\newcommand{\sect}[1]{\setcounter{equation}{0}\section{#1}}
\newcommand{\Red}[1]{\textcolor{red}{#1}}
\renewcommand{\theequation}{\thesection.\arabic{equation}}

\thispagestyle{empty}
\begin{center}
{\Large
{\bf Effects of unparticle on top spin correlation \\
\vspace{2mm}
 at the Large Hadron Collider
}}
\\[8mm]
\vspace{3mm}

\normalsize
{\large \bf
  Masato~Arai~$^{a}$}
\footnote{\it arai@sogang.ac.kr}
,
{\large \bf
  Nobuchika~Okada~$^{b}$}
\footnote{\it
okadan@post.kek.jp}
and 
{\large \bf Karel Smolek~$^{c}$}
\footnote{\it
karel.smolek@utef.cvut.cz
}
\vskip 1.0em

{$^{a}$ \it Center for Quantum Spacetime (CQUeST), Sogang University, \\
  Shinsu-dong 1, Mapo-gu, Seoul 121-742, Korea \\
 $^{b}$ Theory Group, KEK, Tsukuba, 305-0801, Japan \\
 $^{c}$ \it Institute of Experimental and Applied Physics, \\
        Czech Technical University in Prague, 
        Horsk\'a 3a/22, 128 00 Prague 2, Czech Republic
}
\vspace{3mm}
{\bf Abstract}\\[5mm]
{\parbox{16cm}{
We study effects of the scale invariant hidden sector, unparticle, 
 proposed by Georgi, on top spin correlation at the Large Hadron Collider. 
Assuming no flavor changing interaction between the unparticles and the
 Standard Model particles, there arises the top-antitop quark pair production 
 process through virtual unparticle exchanges in the $s$-channel 
 in addition to the Standard Model processes. 
In particular, we consider contributions of scalar and vector
 unparticles and find that these make sizable deviations 
 of the top spin correlation from the Standard Model one. 
}}
\end{center}
\vfill
\newpage
\setcounter{page}{1}
\setcounter{footnote}{0}
\renewcommand{\thefootnote}{\arabic{footnote}}
\section{Introduction}
The Standard Model (SM) quite successfully 
 describes phenomena around the electroweak scale.
However, it is widely believed that new physics beyond the SM appears
 around the TeV scale or higher. 
Recently Georgi proposed a conceptually new possibility that 
 a scale invariant new physics with an infrared fixed point 
 couples to the SM sector \cite{Georgi1,Georgi2}, 
 based on a specific model possessing the scale invariance \cite{Banks}.
Interactions between the new physics sector and the SM sector 
 are realized in the following way.
First we introduce couplings between new physics operator 
 ${\cal O}_{\rm UV}$ with mass dimension $d_{\rm UV}$, 
 which is singlet under the SM gauge group, 
 and the SM operator ${\cal O}_{\rm SM}$ 
 with mass dimension $d_{\rm SM}$ at a mass scale $M$
\begin{eqnarray}
 {\cal L}_{\rm int}={c_n \over M^{d_{\rm UV}+d_{\rm SM}-4}}{\cal O}_{\rm UV}{\cal O}_{\rm SM},
\end{eqnarray}
where $c_n$ is a dimensionless constant.
It is assumed that new physics sector has an infrared fixed point 
 at a scale $\Lambda_{\rm UV}$, below which 
 the operator ${\cal O}_{\rm UV}$ matches onto a new (composite) 
 operator ${\cal O}_{\cal U}$ with dimension $d_{\cal U}$ 
 through the dimensional transmutation. 
As a result, the effective interaction term arises of the form 
\begin{eqnarray}
 {\cal L}_{\rm int}=c_n {\Lambda_{\rm UV}^{d_{\rm UV}-d_{\cal U}} \over M^{d_{\rm
  UV}+d_{\rm SM}-4}}{\cal O}_{\cal U}{\cal O}_{\rm SM}
 \equiv {\lambda_n \over \Lambda^{d_{\cal U}+d_{\rm SM}-4}}{\cal O}_{\cal U}{\cal O}_{\rm
 SM}, \label{int}
\end{eqnarray}
where $\lambda_n$ is a coupling constant and $\Lambda$ is an
 effective cutoff scale of low energy physics.
The operator ${\cal O}_{\cal U}$ is coined as unparticle. 
Depending on the nature of new physics operator ${\cal O}_{\rm UV}$, 
 the resulting unparticle may have different Lorentz structure.
Three unparticle operators, Lorentz scalar ${\cal O}_{\cal U}$, 
 vector ${\cal O}_{\cal U}^\mu$ and tensor ${\cal O}_{\cal U}^{\mu\nu}$,
 were considered  \cite{Georgi1} and their two-point functions were
 derived by the argument based on the scale invariance \cite{Georgi2,ChKeYu1}.
By using them, new phenomena such as direct unparticle emission processes
 \cite{Georgi1} and virtual unparticle exchange processes \cite{Georgi2,ChKeYu1}
 were also discussed.
In particular virtual unparticle exchange process is interesting since
 unparticles with a possible different spin nature affect 
 spin configuration and angular distribution of outgoing SM particles.

A suitable candidate to see effects of virtual unparticles exchange 
 is top spin correlations in the top-antitop pair production process.
The top quark with mass range  of 175 GeV \cite{Abe} decays 
 electroweakly before hadronizing \cite{Bigi},
 and thus the information of polarization of the top-antitop quark pair 
 is directly transferred to its decay products without diluting by hadronization. 
The spin correlations for the hadronic top-antitop pair production 
 process have been extensively studied in the quantum 
 chromodynamics (QCD) \cite{Stelzer, Mahlon-Parke, Bernreuther2}. 
It is then found that there is a spin asymmetry between the produced 
 top-antitop pairs, namely, the number of produced top-antitop quark 
 pairs with both spin up or spin down (like pairs) is different 
 from the number of pairs with the opposite spin combinations (unlike pairs). 
If the top quark is coupled to new physics beyond the SM,
 new physics effects could alter the top-antitop spin correlations. 
Therefore, the top-antitop spin correlations can provide 
 useful information to test not only the SM but also a possible new physics.
Effects of new physics on the top-antitop spin correlations 
 have been studied at the $e^+e^-$ collider \cite{ee} and the photon 
 collider \cite{gamma}.
It should be noticed that the Large Hadron Collider (LHC)
 is a promising laboratory to study the top-antitop quark production
 and the top spin correlations, 
 since it will produce almost 10 millions of top quarks a year. 
Effects of several new physics models
 such as the Kaluza-Klein gravitons in the brane world models 
 \cite{AOSS1, AOSS2} and $Z^\prime$-boson \cite{AOSS3}
 on the top spin correlations at the LHC were studied and 
 sizable deviations of the top spin correlations from the SM 
 one were found.
Also analysis of the top spin correlations
 through possible new physics has been performed 
 in a model independent way with the use of 
 Monte-Carlo simulation \cite{maltoni}.

So far there are some studies of effects of unparticles 
 on top-antitop quark pair production process. 
The total cross section of top-antitop quark pair production 
 through virtual unparticle exchanges was studied 
 at hadron colliders \cite{ChGh}, 
 the International Linear Collider (ILC) \cite{AlPa} and
 the photon collider \cite{LiLiSiYa}. 
For the ILC and the photon collider, the top spin correlation 
 was also studied \cite{Sa1} \cite{Sa2}. 
In this paper we investigate effects of scalar and vector 
 unparticles on the top-antitop pair production and its spin
 correlations at the LHC.
In addition to the SM processes, 
 the unparticle gives rise to a new contribution for the
 top-antitop pair production process  in the $s$-channel 
 through the effective coupling (\ref{int}) 
 and alters the top-antitop pair production cross
 section and the top spin correlations from the SM ones.
 
This paper is organized as follows.
In the next section, we briefly review the top spin correlations.
In section 3, we give a brief review on the basics of unparticle physics. 
In section 4, we derive the invariant amplitudes for the polarized 
 top-antitop pair production processes mediated by the scalar and vector unparticles.
We show the results of our numerical analysis in section 5. 
Section 6 is devoted to conclusions.
Appendix ensembles formulas used in our calculations.

\section{Top spin correlation}
At hadron collider, the top-antitop quark pair is produced 
 through the processes of quark-antiquark pair annihilation 
 and gluon fusion:
\begin{eqnarray}
 &i \rightarrow t+\bar{t}, \,\,\, i=q\bar{q}\,,gg.& \label{top1}
\end{eqnarray}
The former is the dominant process at the Tevatron, 
 while the latter is dominant at the LHC. 
The produced top-antitop pairs decay before hadronization takes place. 
The main decay modes in the SM 
 involve leptonic and hadronic modes: 
\begin{eqnarray}
 &t\rightarrow bW^+ \rightarrow bl^+\nu_l\,,bu\bar{d}\,,bc\bar{s},&
 \label{decay}
\end{eqnarray}
where $l=e,\mu,\tau$.
The differential decay rates to a decay product $f=b,l^+, \nu_l,$ etc.  
 at the top quark rest frame can be parameterized as 
\begin{eqnarray}
{1 \over \Gamma}{d \Gamma \over d \cos \theta_f}=
  {1 \over 2}(1 + \kappa_f \cos \theta_f ), 
 \label{decay1}
\end{eqnarray}
where $\Gamma$ is the partial decay width of 
 the respective decay channel and 
 $\theta_f$ is the angle between the 
 top quark polarization
 and the direction of motion of the decay product $f$.
The coefficient $\kappa_f$ called the top spin analyzing power 
 is a constant between $-1$ and $1$.
The ability to distinguish 
 the polarization
 of the top quark evidently increases with $\kappa_f$.
The most powerful spin analyzer is a charged lepton, for which 
 $\kappa_{l^+}=+1$ at tree level \cite{Jezabek}. 
Other values of $\kappa_f$ are
 $\kappa_b = -0.41$ for the $b$-quark 
 and $\kappa_{\nu_l}=-0.31$ for the $\nu_l$, respectively. 
In hadronic decay modes, the role of the charged lepton 
 is replaced by the $\bar{d}$ or $\bar{s}$ quark.

Now we see how top spin correlations appear in the chain of processes 
 of $i\rightarrow t\bar{t}$ and decay of the top quarks.
The total matrix element squared 
 for the top-antitop pair production \p{top1} 
 and its decay channels \p{decay} is given by 
\begin{eqnarray}
|{\cal M}|^2 \propto {\rm Tr}[\rho R^i \bar{\rho}]
 =\rho_{\alpha^\prime\alpha}R^i_{\alpha\beta,\alpha^\prime\beta^\prime}
  \bar{\rho}_{\beta^\prime\beta} \label{comp}
\end{eqnarray}
in the narrow-width approximation for the top quark.
Here the subscripts denote the top and antitop spin indices, 
 and $R^i$ denotes the density matrix 
 corresponding to the production of the on-shell top-antitop quark pair 
 through the process $i$ in \p{top1}:
\begin{eqnarray}
 R_{\alpha\beta,\alpha^\prime\beta^\prime}^i
 =\sum_{{\rm initial~spin}}{\cal M}(i\rightarrow t_\alpha\bar{t}_\beta)
  {\cal M}^*(i\rightarrow t_{\alpha^\prime}\bar{t}_{\beta^\prime}),
\end{eqnarray}
where ${\cal M}(i\rightarrow t_\alpha\bar{t}_\beta)$ is 
 the amplitude for the top-antitop pair production. 
The matrices $\rho$ and $\bar{\rho}$ are the density matrices 
 corresponding to the decays of polarized top and antitop quarks 
 into some final states at the top and antitop rest frame, respectively.
In the leptonic decay modes, 
 the matrices $\rho$, which lead to \p{decay1},
 can be obtained as (see, for instance, \cite{Bernreuther})
\begin{eqnarray}
  \rho_{\alpha^\prime\alpha}
  = {\cal M}(t_\alpha \rightarrow bl^+\nu_l)
    {\cal M}^*(t_{\alpha^\prime} \rightarrow bl^+\nu_l)  
  = {\Gamma \over 2}(1 + \kappa_f {\vec{\sigma}} \cdot 
       \vec{q}_f)_{\alpha^\prime\alpha},  
 \label{rho1}
\end{eqnarray}
where $q_f$ is the unit vector of the direction of motion 
 of the decay product $f$. 
The density matrix for the polarized antitop quark 
 is obtained by replacing $\kappa_f \rightarrow -\kappa_f$ in \p{rho1}
 if there is no CP violation.
In the SM, there is no CP violation in the top quark decay at the
 leading order. 
In the model presented in the next section,
 there is no contribution to break CP symmetry at the leading order, and
 thus this relation holds.

A way to analyze the top-antitop spin correlations 
 is to see the angular correlations 
 of two charged leptons $l^+l^-$ 
 produced by the top-antitop quark leptonic decays. 
In the following, we consider only the leptonic decay channels. 
Using \p{comp}-\p{rho1} and integrating over 
 the azimuthal angles of the charged leptons, 
 we obtain the following double distribution
 \cite{Stelzer, Mahlon-Parke}
\begin{eqnarray}
 {1 \over \sigma}
 {d^2 \sigma \over d \cos\theta_{l^+} d \cos\theta_{l^-}}
 = {1 \over 4}\left({1+B_1 \cos\theta_{l^+}
 +B_2 \cos\theta_{l^-}-C
    \cos\theta_{l^+} \cos\theta_{l^-}}\right).
 \label{double}
\end{eqnarray}
Here $\sigma$ denotes the cross section 
 for the process of the leptonic decay modes, 
 and $\theta_{l^+} (\theta_{l^-})$ denotes the angle 
 between the top (antitop) spin axis and 
 the direction of motion of the antilepton (lepton) 
 at the top (antitop) rest frame. 
In what follows, we use the helicity spin basis which is almost optimal
 one to analyze the top spin correlation at the LHC\footnote{
 See \cite{uwer} for the study on another spin basis, 
 which has a larger spin correlation than the helicity basis 
 at the LHC.}. 
In this basis, the top (antitop)
 spin axis is regarded as the direction of motion of the top (antitop) in
 the top-antitop center-of-mass system.
The coefficients $B_1$ and $B_2$ are associated with a polarization of the top
 and antitop quarks, 
 and $C$ encodes the top spin correlations,
 whose explicit expression is given by
\begin{eqnarray}
 C= {\cal A} \kappa_{l^+}\kappa_{l^-},~~~\kappa_{l^+}=\kappa_{l^-}=1,
\end{eqnarray}
where the coefficient ${\cal A}$ represents the spin asymmetry 
 between the produced top-antitop pairs 
 with like and unlike spin pairs defined as 
\begin{eqnarray}
 {\cal A}={\sigma(t_\uparrow\bar{t}_\uparrow)
          +\sigma(t_\downarrow\bar{t}_\downarrow) 
          -\sigma(t_\uparrow\bar{t}_\downarrow)
          -\sigma(t_\downarrow\bar{t}_\uparrow) 
         \over
           \sigma(t_\uparrow\bar{t}_\uparrow)
          +\sigma(t_\downarrow\bar{t}_\downarrow)
          +\sigma(t_\uparrow\bar{t}_\downarrow)
          +\sigma(t_\downarrow\bar{t}_\uparrow)}. 
\label{asym}
\end{eqnarray}
Here $\sigma(t_\alpha\bar{t}_\beta)$ is 
 the cross section of the top-antitop pair production 
 at parton level with denoted spin indices.

In the SM, at the lowest order of $\alpha_s$, 
 the spin asymmetry is found to be ${\cal A}=+0.319$ 
 for the LHC\footnote{
The parton distribution function set of CTEQ6L \cite{CTEQ} 
 has been used in our calculations. 
The resultant spin asymmetry somewhat depends 
 on the parton distribution functions used.}. 
At the LHC in the ATLAS experiment, the spin asymmetry 
 of the top-antitop pairs will be 
 measured with a precision of several percent, after one LHC year 
 at low luminosity (10 fb$^{-1}$) \cite{ATLAS_TOP}. 
This accuracy can enhance  the feasibility to find new physics 
 effects at the LHC through the top spin correlation.

\section{Unparticle physics}
We briefly review derivations of two-point functions 
 of scalar and vector unparticles, which are relevant for our analysis.
It was argued in \cite{Georgi2} that the scale invariance can be used to
 fix the two-point function of unparticle operators
\begin{eqnarray}
 \langle 0|{\cal O}_{\cal U}(x){\cal O}_{\cal U}^\dagger(0)| 0 \rangle
 =\int {d^4 P \over (2\pi)^2}e^{-iP\cdot x}\rho(P^2), \label{op1}
\end{eqnarray}
where $\rho(P^2)=(2\pi)^2\int d\lambda 
 \delta^4(P-p_\lambda)|\langle 0|{\cal O}_{\cal U}|\lambda \rangle|^2$.
The spectral function $\rho(P^2)$ is determined by scale invariance to be
 $\rho(P^2)=A_{d_{\cal U}}\theta(P^0)\theta(P^2)(P^2)^{d_{\cal U}-2}$,
 where $A_{d_{\cal U}}$ is the normalization factor.
This factor is fixed by identifying $\rho(P^2)$ with
 $d_{\cal U}$-body phase space of massless particle to be
\begin{eqnarray}
 A_{d_{\cal U}}={16\pi^2 \sqrt{\pi} \over (2\pi)^{2d_{\cal U}}}
 {\Gamma(d_{\cal U}+1/2) \over \Gamma(d_{\cal U}-1)\Gamma(2d_{\cal U})}. 
\end{eqnarray}
With the use of the spectral function $\rho(P^2)$ and requiring scale
 invariance, we can define the Feynman propagator.
The propagator for the scalar unparticle is given by \cite{Georgi2}
\begin{eqnarray}
 \Delta(p)={iA_{d_{\cal U}} \over 2\sin(d_{\cal U}\pi)}(-p^2)^{d_{\cal U}-2},
\end{eqnarray}
and similarly for the vector unparticle (with only the transverse mode)
\begin{eqnarray}
 \Delta^{\mu\nu}(p)={iA_{d_{\cal U}} \over
  2\sin(d_{\cal U}\pi)}(-p^2)^{d_{\cal U}-2}
  \left(g^{\mu\nu}-{p^\mu p^\nu \over p^2}\right).\label{vec-p2}
\end{eqnarray}

We could also consider the rigid conformal invariance as a symmetry of
 hidden sector \cite{Mack,Grinstein}.
By requiring conformal invariance, the scalar unparticle propagator
 remains the same form while the vector unparticle propagator is
 modified to 
\begin{eqnarray}
  \Delta^{\mu\nu}(p)={iA_{d_{\cal U}} \over
  2\sin(d_{\cal U}\pi)}(-p^2)^{d_{\cal U}-2}
  \left(g^{\mu\nu}-{2(d_{\cal U}-2) \over d_{\cal U}-1}{p^\mu p^\nu \over p^2}\right).
\end{eqnarray}
In Ref. \cite{Grinstein}, the theoretical bound of the scaling dimension was
 obtained from unitarity argument in this case.
The scaling dimension for the scalar unparticle is constrained as
 $d_{\cal U}\ge 1$ while for the vector unparticle the bound
 is $d_{\cal U}\ge 3$.
The vector unparticle interaction with the latter bound is very suppressed
 and it would not cause sizable deviation from the SM.
In this paper, we will concentrate on the scale invariant hidden 
 sector, but we will also show some results for the conformal invariant
 hidden sector with the scaling dimension $d_{\cal U}=3.01$, 
 satisfying the above-mentioned bound (see Table 1)\footnote{
In analysis of the top spin correlations, 
 the term proportional to $p^\mu p^\nu$ in the vector unparticle 
 propagator is vanishing under the equation of motion 
 for the initial (almost massless) light quark. 
 Therefore, the difference between the scale invariant and 
 the conformal invariant theories is just the bound for
 the scaling dimension.}.

In the following we list operators composed of SM fields and derivatives
 which are relevant in our consideration.
Relevant effective interactions of the scalar unparticle with 
 the SM fields are given by, for gluon
\begin{eqnarray}
 {\lambda_{gg} \over \Lambda^{d_{\cal U}}}
 {\rm tr}(G^{\mu\nu}G_{\mu\nu}){\cal O}_{\cal U}, \label{gluon}
\end{eqnarray}
where $\lambda_{gg}$ is a constant.
For fermions we have (up to dimensionless coefficients), 
\begin{eqnarray}
&&{1 \over \Lambda^{d_{\cal U}}}\bar{Q}_L\gamma^\mu Q_L\partial_\mu {\cal
 O}_{\cal U},~~
{1 \over \Lambda^{d_{\cal U}}}\bar{U}_R\gamma^\mu U_R\partial_\mu {\cal
 O}_{\cal U},~~
{1 \over \Lambda^{d_{\cal U}}}\bar{D}_R\gamma^\mu D_R\partial_\mu {\cal
 O}_{\cal U},~~\label{s1} \\
&&{1 \over \Lambda^{d_{\cal U}}}\bar{Q}_L\gamma^\mu D_\mu Q_L{\cal O}_{\cal U},~~
{1 \over \Lambda^{d_{\cal U}}}\bar{U}_R\gamma^\mu D_\mu U_R {\cal O}_{\cal U},~~
{1 \over \Lambda^{d_{\cal U}}}\bar{D}_R\gamma^\mu D_\mu D_R {\cal O}_{\cal U},~~\label{s2}
\end{eqnarray}
where $Q_L$ is a left-handed quark, and
 $U_R(D_R)$ denotes a right-handed up(down)-type quark.
The interactions with fermions can be simplified by utilizing 
 the equation of motion for a fermion 
\begin{eqnarray}
 i\gamma^\mu \partial_\mu \psi=m_f \psi,
\end{eqnarray}
where $m_f$ is a fermion mass.
Consequently, (\ref{s1}) and (\ref{s2}) are summarized to be
\begin{eqnarray}
 &&{m_Q \over \Lambda^{d_{\cal U}}}\bar{Q}(a_Q^S+i\gamma^5 b_Q^S)Q,\label{scalar-int}
\end{eqnarray}
 where $Q=U,D$ are mass eigenstates of quarks, 
 and $a_Q^S$ and $b_Q^S$ are constants.

Possible interacting terms with the vector unparticle are 
\begin{eqnarray}
&&{1 \over \Lambda^{d_{\cal U}-1}}\bar{Q}_L\gamma^\mu Q_L({\cal O}_{\cal U})_\mu,~~
{1 \over \Lambda^{d_{\cal U}-1}}\bar{U}_R\gamma^\mu U_R({\cal O}_{\cal U})_\mu,~~
{1 \over \Lambda^{d_{\cal U}-1}}\bar{D}_R\gamma^\mu D_R({\cal O}_{\cal U})_\mu.~~\label{v1}
\end{eqnarray}
They are also simplified as
\begin{eqnarray}
 {1 \over \Lambda^{d_{\cal U}-1}}\bar{Q}\gamma^\mu(c_{L}^{Q}P_L + c_R^{Q}P_R)
  Q({\cal O}_{\cal U})_\mu, \label{vector-int}
\end{eqnarray}
where $c_L^Q$ and $c_R^Q$ are coupling constants.
%
%
\section{Amplitudes}
In this section we calculate the squared invariant amplitudes for
 $q\bar{q}\rightarrow t\bar{t}$ and $gg\rightarrow t\bar{t}$ processes.
First we consider effect of the scalar unparticle. 
In this case, we only consider the $gg\rightarrow t\bar{t}$
 process for new contribution by the scalar unparticle, 
 because the $q\bar{q}\rightarrow t\bar{t}$ process is 
 proportional to light quark mass and hence negligible. 

Since there is no interference between the QCD and 
 the scalar unparticle mediated processes, 
 the squared amplitude for the $gg\rightarrow t\bar{t}$ process 
 is simply given by 
\begin{eqnarray}
 |{\cal M}(gg\rightarrow t\bar{t})|^2
  =|{\cal M}_{\rm QCD}(gg\rightarrow t\bar{t})|^2
  +|{\cal M}_{\rm SU}(gg\rightarrow t\bar{t})|^2, 
\end{eqnarray}
where ${\cal M}_{\rm QCD}$ is the amplitude of the QCD process  
 and ${\cal M}_{\rm SU}$ is the contribution of the scalar unparticle.
We calculate the helicity decomposition of the above amplitude with
 respect to the final top spin polarization.
For the squared amplitude for the QCD process with the $gg$ initial
 state, we have
\begin{eqnarray} 
 |{\cal M}_{\rm QCD}(gg \rightarrow t_\uparrow\bar{t}_\uparrow)|^2
    &=& |{\cal M}_{\rm QCD}(gg \rightarrow
    {t_\downarrow\bar{t}_\downarrow})|^2\nonumber \\
 &=&{g_s^4 \over 96}{\cal Y}(\beta_t,\cos\theta)(1-\beta_t^2)
     (1+\beta_t^2+\beta_t^2\sin^4\theta),\label{gg1}\\
 |{\cal M}_{\rm QCD}(gg\rightarrow t_\uparrow\bar{t}_\downarrow)|^2
   &=& |{\cal M}_{\rm QCD}(gg\rightarrow {t_\downarrow\bar{t}_\uparrow})|^2
   ={g_s^4 \beta_t^2 \over 96}{\cal
     Y}(\beta_t,\cos\theta)\sin^2\theta(1+\cos^2\theta), \label{gg2}
\end{eqnarray}
where $g_s$ is the strong coupling constant, 
 $\beta_t=\sqrt{1-4m_t^2/s}$, $m_t$ is the top quark mass, 
 $\sqrt{s}$ is energy of colliding partons
 and $\theta$ is the scattering angle between incoming quark and outgoing
 top quark.
The form of ${\cal Y}(\beta_t,\theta)$ is defined by
\begin{eqnarray}
 {\cal Y}(\beta_t,\cos\theta)&=&
  {7+9\beta_t^2\cos^2 \theta \over (1-\beta_t^2\cos^2\theta)^2}.
  \label{funcs1}
\end{eqnarray}
The squared helicity amplitude mediated by scalar unparticle is written by
\begin{eqnarray}
 |{\cal M}_{\rm SU}(gg\rightarrow t_\gamma\bar{t}_\delta)|^2
 =\left({1 \over 2}\right)^2
 \left({1 \over 3^2-1}\right)^2{(3^2-1)3 \over 4}
 \sum_{\lambda_1,\lambda_2}
 |{\cal M}_{\rm SU}(\lambda_1,\lambda_2;\gamma,\delta)|^2,
\end{eqnarray}
where $\lambda_i(i=1,2)=\pm 1$ are the initial spins of gluons,
 $\gamma=\pm (\delta=\pm)$ denotes spin up/down for the final state 
 top (antitop) quark. 
The amplitude
 ${\cal M}_{\rm SU}(\lambda_1\lambda_2\rightarrow t_\gamma\bar{t}_\delta)$ 
 is the helicity decomposition of 
 ${\cal M}_{\rm SU}(gg\rightarrow t_\gamma\bar{t}_\delta)$
 with respect to the initial spins, given by
\begin{eqnarray}
&& {\cal M}_{\rm SU}(\lambda_1,\lambda_2,\pm,\pm)
 =\pm{A_{d_{\cal U}}\lambda_{gg}m_t e^{i\pi(d_{\cal U}-1/2)} \over \sin(d_{\cal U} \pi)
  \Lambda^{2d_{\cal U}}}s^{d_{\cal U}-1/2}{1+\lambda_1\lambda_2 \over 2}
  (a_t^S\beta_t\mp i b_t^S),\label{su1}\\
&& {\cal M}_{\rm SU}(\lambda_1,\lambda_2,\pm,\mp)=0.\label{su2}
\end{eqnarray}

For the $q\bar{q}\rightarrow t\bar{t}$ process, we have
\begin{eqnarray}
|{\cal M}(q\bar{q}\rightarrow t\bar{t})|^2=
 |{\cal M}_{\rm QCD}(q\bar{q} \rightarrow t\bar{t})|^2
 +|{\cal M}_{\rm NC}(q\bar{q} \rightarrow t\bar{t})|^2,
\end{eqnarray}
where 
${\cal M}_{\rm QCD}$ and ${\cal M}_{\rm NC}$ are the amplitudes 
 of the QCD and the neutral current processes, respectively.
The helicity decomposition of ${\cal M}_{\rm QCD}$ 
 with respect to the final state is given by
\begin{eqnarray}
 |{\cal M}_{\rm QCD}(q\bar{q}\rightarrow t_\uparrow\bar{t}_\uparrow)|^2
  &=& |{\cal M}_{\rm QCD}(q\bar{q}\rightarrow {t_\downarrow\bar{t}_\downarrow})|^2
  = {g_s^4 \over 9}(1-\beta_t^2)\sin^2\theta,
  \label{qq1} \\
 |{\cal M}_{\rm QCD}(q\bar{q}\rightarrow t_\uparrow\bar{t}_\downarrow)|^2
  &=& |{\cal M}_{\rm QCD}(q\bar{q}\rightarrow {t_\downarrow\bar{t}_\uparrow})|^2
  = {g_s^4 \over 9}(1+\cos^2\theta).
  \label{qq2}
\end{eqnarray}
The helicity amplitude of ${\cal M}_{\rm NC}$ is written as
\begin{eqnarray}
 |{\cal M}_{\rm NC}(q\bar{q} \rightarrow t_\gamma\bar{t}_\delta)|^2
 =\left({1 \over 2}\right)^2\sum_{\alpha,\beta}
  |{\cal M}_{\rm NC}(\alpha,\beta;\gamma,\delta)|^2,
\end{eqnarray}
where ${\cal M}_{\rm NC}(\alpha,\beta;\gamma,\delta)$ are
 the helicity amplitudes and the symbols $\alpha(\gamma)$ 
 and $\beta(\delta)$ denote initial (final) 
 spin states for quark and antiquark, respectively.
They are described by (color factor is suppressed)
\begin{eqnarray}
&&\hspace{-10mm}{\cal M}_{\rm NC}(+,-;\pm,\pm)=\mp s\sqrt{1-\beta_t^2}\sin\theta
  \left[{(eQ^f)(eQ^t) \over s}+
  {g_{R}^f \over 2}{g_{L}^t+g_{R}^t \over
  s-M_{Z}^2+iM_{Z}\Gamma_{Z}} \right],\\
&&\hspace{-10mm}{\cal M}_{\rm NC}(-,+;\pm,\pm)=\mp s\sqrt{1-\beta_t^2}\sin\theta
  \left[{(eQ^f)(eQ^t) \over s}+
  {g_{L}^f \over 2}{g_{L}^t+g_{R}^t \over
  s-M_{Z}^2+iM_{Z}\Gamma_{Z}} \right], \\
&&\hspace{-10mm}{\cal M}_{\rm NC}(+,-;+,-)=-s(1+\cos\theta)\left[
  {(eQ^f)(eQ^t) \over s}+{g_{R}^f \over 2}
 {(1-\beta_t)g_{L}^t+(1+\beta_t)g_{R}^t \over s-M_{Z}^2+iM_{Z}\Gamma_{Z}}
  \right],\\
&&\hspace{-10mm}{\cal M}_{\rm NC}(+,-;-,+)=s(1-\cos\theta)\left[
  {(eQ^f)(eQ^t)\over s}+{g_{R}^f \over 2}
 {(1+\beta_t)g_{L}^t+(1-\beta_t)g_{R}^t \over s-M_{Z}^2+iM_{Z}\Gamma_{Z}}
  \right],\\
&&\hspace{-10mm}{\cal M}_{\rm NC}(-,+;+,-)=s(1-\cos\theta)\left[
  {(eQ^f)(eQ^t) \over s} + {g_{L}^f \over 2}
 {(1-\beta_t)g_{L}^t+(1+\beta_t)g_{R}^t \over s-M_{Z}^2+iM_{Z}\Gamma_{Z}}
  \right],\\
&&\hspace{-10mm}{\cal M}_{\rm NC}(-,+;-,+)=-s(1+\cos\theta)\left[
  {(eQ^f)(eQ^t) \over s} + {g_{L}^f \over 2}
 {(1+\beta_t)g_{L}^t+(1-\beta_t)g_{R}^t \over s-M_{Z}^2+iM_{Z}\Gamma_{Z}}
  \right], \label{gamma}
\end{eqnarray}
with the decay widths of $Z$ boson $\Gamma_Z$ given by
\begin{eqnarray}
 \Gamma_{Z}=\Gamma(Z\rightarrow f\bar{f})
 ={M_{Z}\over 96\pi}\sum_{f}\beta^f
  \left\{(3+(\beta^{f})^2)((g_{L}^f)^2+(g_{R}^f)^2)+6(1-(\beta^f)^2)g_{L}^f
   g_{R}^f\right\}.\label{gamma-total}
\end{eqnarray}
Here $M_{Z}$ is the mass of the $Z$-boson and
 $\beta^f=\sqrt{1-4m_f^2/M_{Z}^2}$.
Couplings, charges and the decay widths $\Gamma_Z$ are explicitly given in
 Appendix \ref{appendix-A}.

Next we calculate the case for the vector unparticle.
It contributes to the quark annihilation process 
 $q\bar{q} \rightarrow  t\bar{t}$ in the $s,t$ and $u$-channels
 in addition to the Standard Model processes.
In our analysis, we assume no flavor violating processes and therefore
 we only consider the vector unparticle exchange in $s$-channel process.
A total amplitude for quark annihilation process is given by 
\begin{eqnarray}
 {\cal M}(q\bar{q}\rightarrow t\bar{t})&=&
 {\cal M}_{\rm NC}(q\bar{q}\rightarrow t\bar{t})
  +{\cal M}_{\rm QCD}(q\bar{q}\rightarrow t\bar{t})
 +{\cal M}_{\rm VU}(q\bar{q}\rightarrow t\bar{t}),
\end{eqnarray}
where ${\cal M}_{\rm NC}$ is the neutral current process,
 ${\cal M}_{\rm QCD}$ is the QCD process given in (\ref{qq1}) and (\ref{qq2}),
 and ${\cal M}_{\rm VU}$ is the contribution of the vector unparticle.
Since there is no interference between the QCD process
 and other processes,
 the squared amplitude is written as 
\begin{eqnarray}
 |{\cal M}(q\bar{q}\rightarrow t\bar{t})|^2
  &=&|({\cal M}_{\rm NC}
   +{\cal M}_{\rm VU})(q\bar{q}\rightarrow t\bar{t})|^2
   +|{\cal M}_{\rm QCD}(q\bar{q}\rightarrow t\bar{t})|^2.
\end{eqnarray}
The helicity amplitude of the neutral current process and
 the vector unparticle mediated process is described by 
\begin{eqnarray}
&&|({\cal M}_{\rm NC}
   +{\cal M}_{\rm VU})(q\bar{q}\rightarrow t_\gamma\bar{t}_\delta)|^2 \nonumber \\
&&~~ =\left({1 \over 2}\right)^2
  \sum_{\alpha,\beta}\left(|{\cal M}_{\rm NC}(\alpha,\beta;\gamma,\delta)|^2
  +|{\cal M}_{\rm VU}(\alpha,\beta;\gamma,\delta)|^2
 +({\cal M}_{\rm NC}{\cal M}_{\rm
 VU}^*(\alpha,\beta;\gamma,\delta)+h.c.)\right),\nonumber \\
\end{eqnarray}
where ${\cal M}_{\rm VU}(\alpha,\beta;\gamma,\delta)$ are the decompositions
 of the helicity amplitudes 
 ${\cal M}_{\rm VU}(q\bar{q}\rightarrow t_\gamma\bar{t}_\delta)$  
 with respect to the initial spin.

The helicity amplitudes mediated by the vector unparticle 
 ${\cal M}_{\rm VU}(\alpha,\beta;\gamma,\delta)$ are given by 
\begin{eqnarray}
{\cal M}_{\rm VU}(+,-;\pm,\pm)&=& 
 \pm s^{d_{\cal U}-1}\sqrt{1-\beta_t^2}\sin\theta
  {A_{d_{\cal U}} e^{i\pi(d_{\cal U}-2)} \over 2\sin(d_{\cal U}\pi)\Lambda^{2(d_{\cal U}-1)}}{c_R^Q
  \over 2}(c_L^t+c_R^t),\label{VU1} \\
{\cal M}_{\rm VU}(-,+;\pm,\pm)&=& 
  \pm s^{d_{\cal U}-1}\sqrt{1-\beta_t^2}\sin\theta
  {A_{d_{\cal U}} e^{i\pi(d_{\cal U}-2)} \over 2\sin(d_{\cal U}\pi)\Lambda^{2(d_{\cal U}-1)}}{c_L^Q
  \over 2}(c_L^t+c_R^t),\label{VU2} \\
{\cal M}_{\rm VU}(+,-;\pm,\mp)&=& 
 s^{d_{\cal U}-1}(\cos\theta\pm 1)
  {A_{d_{\cal U}} e^{i\pi(d_{\cal U}-2)} \over 2\sin(d_{\cal U}\pi)\Lambda^{2(d_{\cal U}-1)}}
 {c_R^Q \over 2}\left(c_L^t+ c_R^t \mp \beta_t(c_L^t - c_R^t)\right),
 \label{VU3} \\
{\cal M}_{\rm VU}(-,+;\pm,\mp)&=& 
 s^{d_{\cal U}-1}(\cos\theta\mp 1)
  {A_{d_{\cal U}} e^{i\pi(d_{\cal U}-2)} \over 2\sin(d_{\cal U}\pi)\Lambda^{2(d_{\cal U}-1)}}
 {c_L^Q \over 2}\left(c_L^t+ c_R^t \mp \beta_t(c_L^t-
		 c_R^t)\right),\label{VU4}
\end{eqnarray}
where $c_L^t(c_R^t)$ is the coupling constant $c_L^Q(c_R^Q)$ in (\ref{vector-int})
with $Q=t$.

\section{Numerical analysis}
Here we show various numerical results 
 and demonstrate interesting properties of measurable quantities. 
In our analysis we use the parton distribution function of 
 CTEQ6L \cite{CTEQ} with the factorization scale $Q=m_t=175$
 GeV and $\alpha_s(Q)=0.1074$. 
In the whole analysis, the center of mass energy of the colliding
 protons, $E_{CMS}$, is taken to be $1.96$ TeV at the Tevatron and $14$ TeV at the LHC.
For simplicity, we fix model parameters as follows: 
 $\lambda_{gg}=1$ in (\ref{gluon}), 
 $a_Q^S=b_Q^S=c_L^Q=c_R^Q=1$ 
 in (\ref{scalar-int}) and (\ref{vector-int}), 
 and $\Lambda=1$ TeV. 

\begin{figure}[ht]
\begin{center}
  \epsfxsize=12cm
  \epsfbox{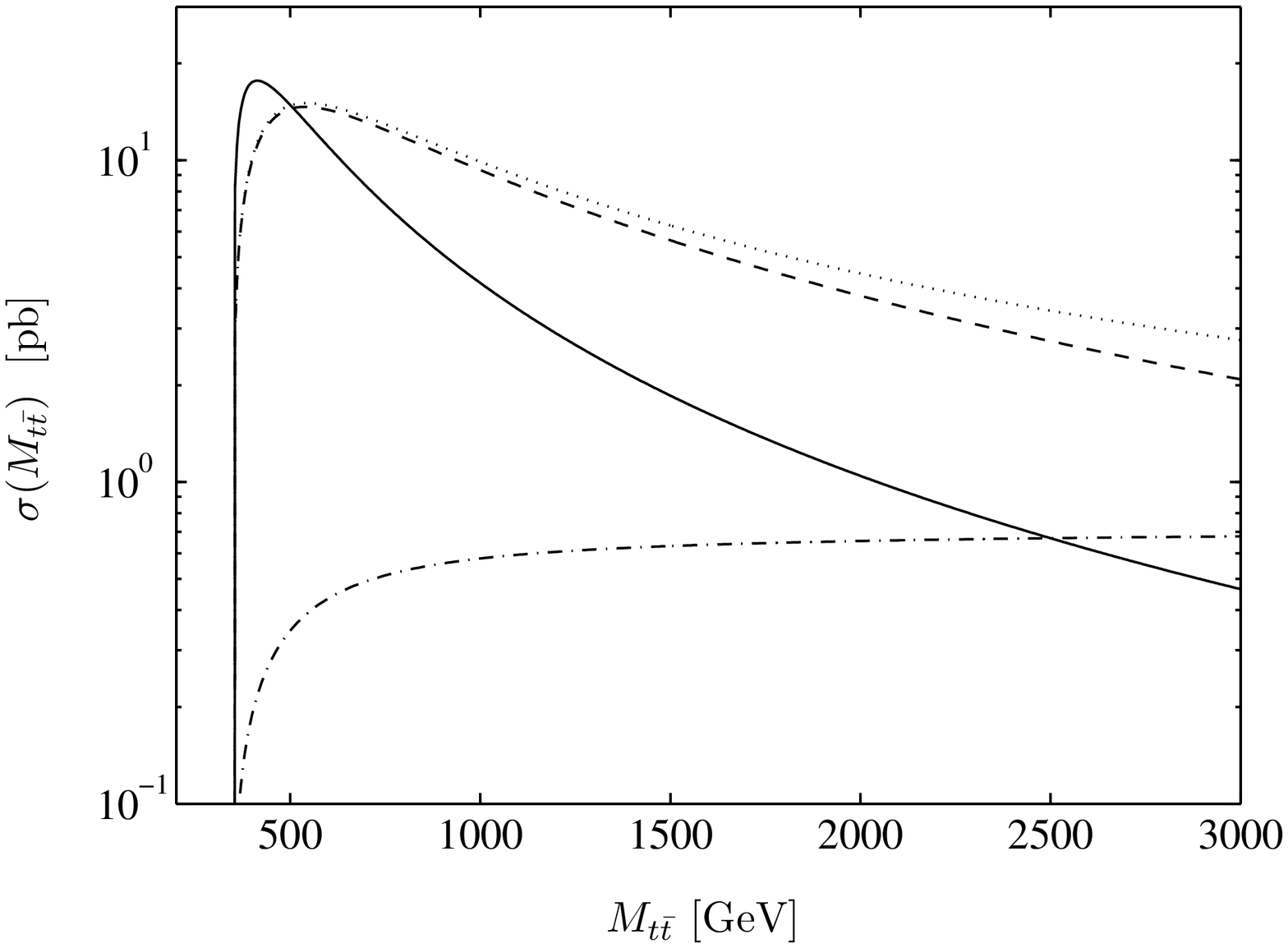}
\caption{The dependence of the  
 cross section of the top-antitop 
 quark pair production by quark pair annihilation and gluon fusion
 on the center-of-mass energy of colliding partons $M_{t\bar{t}}$ 
 with $d_{\cal U}=1.01$ and $\Lambda=1$ TeV. 
The solid and dashed lines correspond to 
 the results of up-quark annihilation and gluon fusion for the SM, respectively.
The dotted and dash-dotted lines correspond to the results of
 the SM (gluon fusion) $+$ the scalar unparticle processes and only the scalar unparticle
 contribution.}
\label{fig_cross_parton_SU}
\end{center}
\end{figure}

\begin{figure}[ht]
\begin{center}
  \epsfxsize=12cm
  \epsfbox{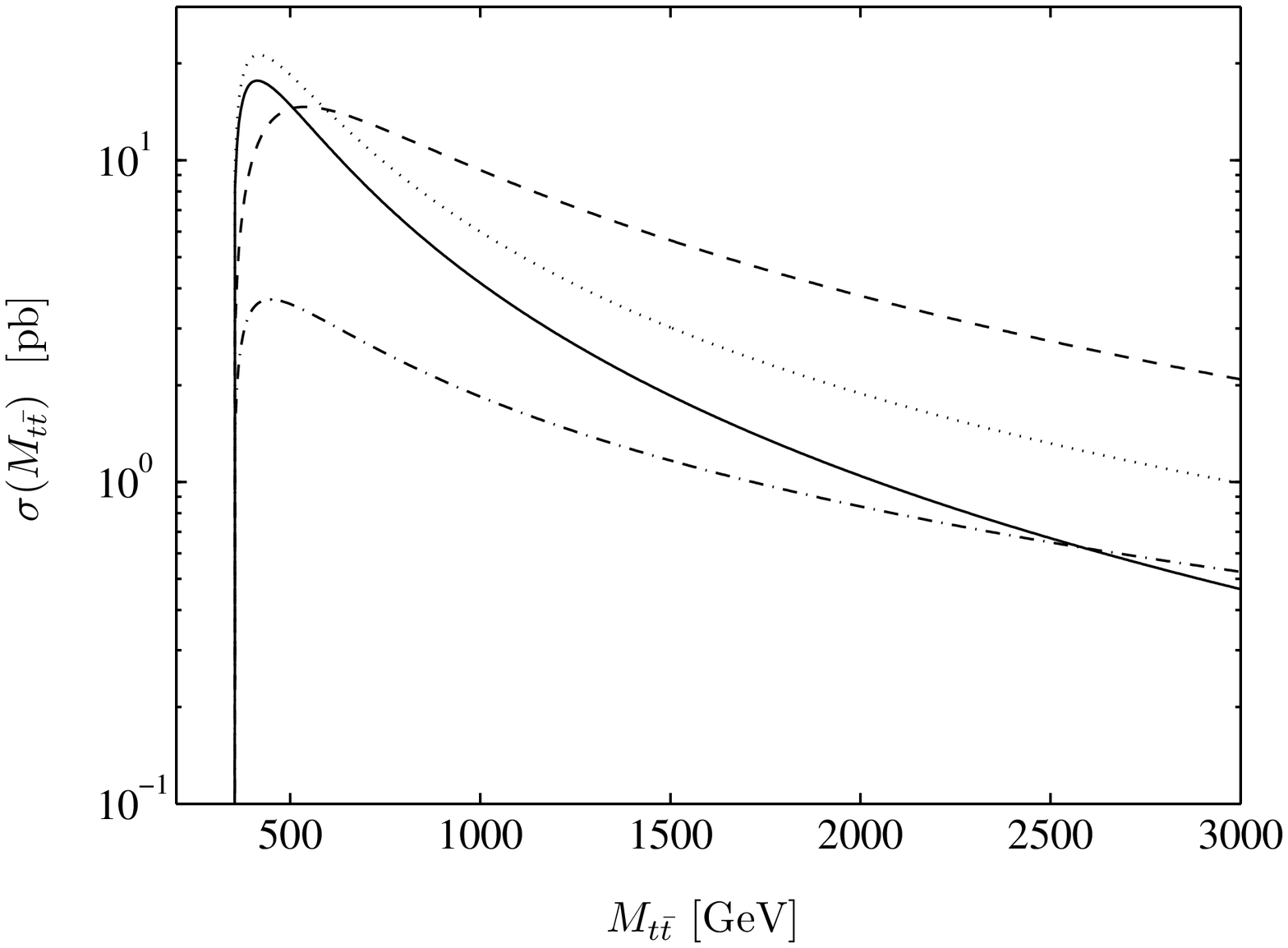}
\caption{The dependence of the  
 cross section of the top-antitop 
 quark pair production by quark pair annihilation and gluon fusion
 on the center-of-mass energy of colliding partons $M_{t\bar{t}}$
 with $d_{\cal U}=1.01$ and $\Lambda=1$ TeV. 
The solid and dashed lines correspond to 
 the results of the up-quark annihilation and gluon fusion for the SM, respectively.
The dotted and dash-dotted lines correspond to the results of
 the SM (up-quark annihilation) $+$ vector unparticle processes and 
 only the vector unparticle contribution.}
\label{fig_cross_parton_VU}
\end{center}
\end{figure}

As can be seen from the formulas of the squared
 amplitudes (\ref{su1}), (\ref{VU1})--(\ref{VU4}),
 the cross sections through the unparticle exchange processes
 grow or slowly decrease compared the SM cross sections 
 according to the colliding partons center-of-mass energy. 
 When the cross section grows as a power of the center-of-mass energy, 
 the unitarity will be violated at high energies. 
This behavior is shown, for instance, in Figs. \ref{fig_cross_parton_SU} 
 and \ref{fig_cross_parton_VU},  
 where the cross sections of the top-antitop pair production through 
 $q\bar{q}\rightarrow t\bar{t}$ and $gg\rightarrow t\bar{t}$
 at the parton level, respectively, are depicted as a function of
 partons center-of-mass energy $M_{t\bar{t}}$.
In order to make our analysis conservative, 
 we take into account the contributions from unparticle exchange processes
 only for the center-of-mass energy of colliding 
 partons lower than $\Lambda$, namely $\sqrt{s}=M_{t\bar{t}}\le\Lambda$.

\begin{figure}[ht]
\begin{center}
  \epsfxsize=12cm
  \epsfbox{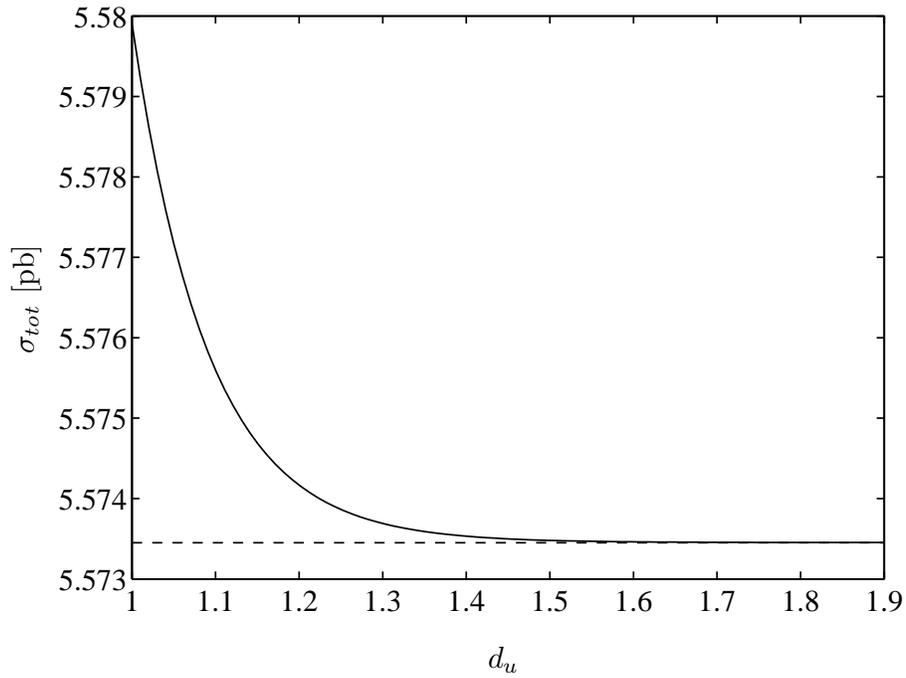}
\caption{The total cross section of the top-antitop 
 quark pair production as a function of $d_{\cal U}$ 
 at Tevatron with $\sqrt{s} =1.96$ TeV and $\Lambda=1$ TeV. 
The solid curve shows the value of the SM + scalar unparticle while the
 dashed line shows the SM value.
}
\label{fig_cross_Tevatron_SU}
\end{center}
\end{figure}

\begin{figure}[ht]
\begin{center}
  \epsfxsize=12cm
  \epsfbox{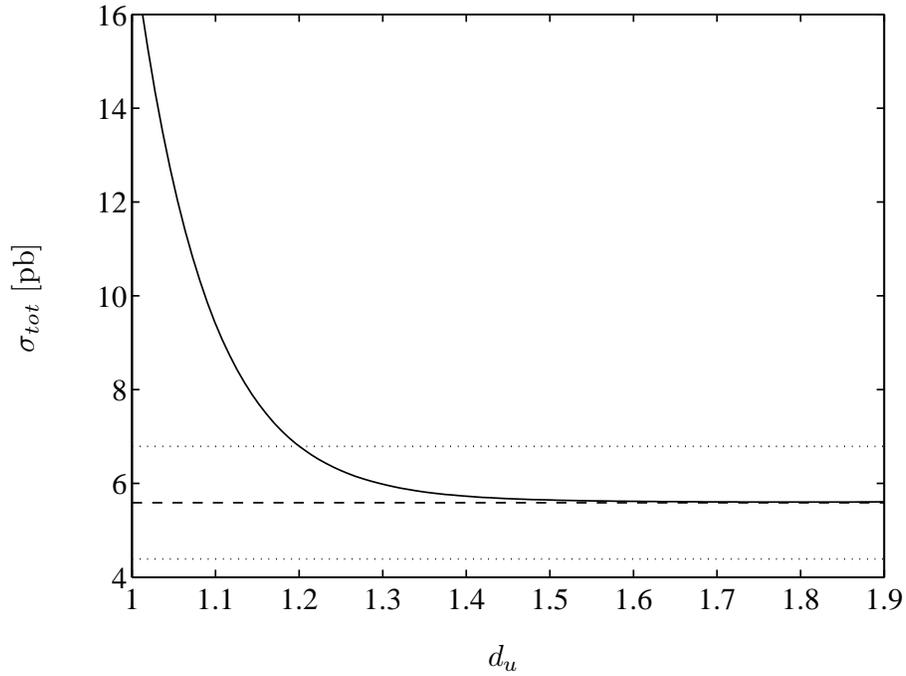}
\caption{The total cross section of the top-antitop 
 quark pair production as a function of $d_{\cal U}$ 
  at Tevatron with $\sqrt{s} =1.96$ TeV and $\Lambda=1$ TeV. 
The solid curve shows the contribution of the vector unparticle to the total cross
 section.
The dashed line corresponds to the SM value,
 and the dotted lines correspond to the estimated errors 
 from Tevatron measurement.
}
\label{fig_cross_Tevatron_VU}
\end{center}
\end{figure}

The scaling dimension of the unparticle $d_{\cal U}$ 
 is a unique free parameter in our analysis. 
Since the Tevatron results for the total cross section of 
 the top-antitop production are consistent 
 with the SM prediction \cite{CDF_cross}, we can obtain 
 the lower bounds for $d_{\cal U}$ from the Tevatron results. 
In Fig. \ref{fig_cross_Tevatron_SU} and \ref{fig_cross_Tevatron_VU}, 
 we present the dependence of the total cross section 
 on $d_{\cal U}$ in the case of the scalar and the vector unparticles. 
The solid line corresponds to the model with unparticle while
 the dashed line corresponds to the SM value. 
We find the SM cross section for the top-antitop pair production 
 at the leading order (LO) as $\sim 5.57$ pb, 
 while in the next-to-next LO (NNLO) analysis the SM prediction 
 is found to be $6.7^{+0.7}_{-0.9}$ pb \cite{Cacciari}. 
Scaling our results to the NNLO value, 
 we estimate the error of the Tevatron measurement as $\pm1.2$ pb
 and apply this error bar to obtain the lower bound on $d_{\cal U}$ 
 (see Fig. \ref{fig_cross_Tevatron_VU}). 
From the plots, we find that there is no bound on 
 $d_{\cal U} (\geq 1)$ for the scalar unparticle 
 while $d_{\cal U} \geq 1.2$ for the vector unparticle.

\begin{figure}[ht]
\begin{center}
  \epsfxsize=12cm
  \epsfbox{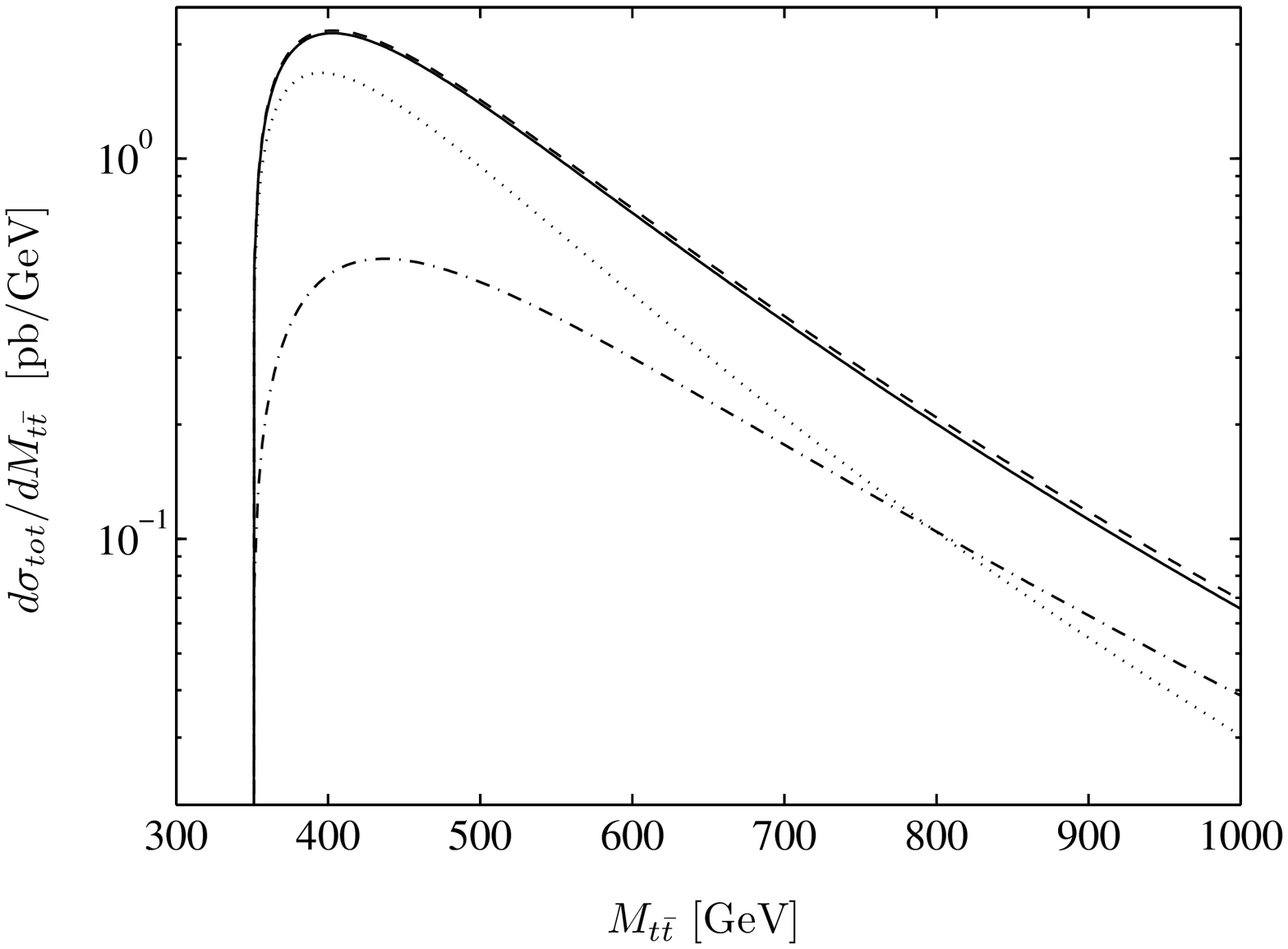}
\caption{Differential cross section \p{total_s} as a function 
 of the top-antitop invariant mass $M_{t\bar{t}}$ for the SM $+$ scalar 
 unparticle processes 
 with $d_{\cal U}=1.01$ and $\Lambda=1$ TeV. 
The solid line depicts the result of the SM and the 
 dashed line depicts the result of the SM $+$ scalar unparticle.
Breakdown of the latter into the like (dotted) and the unlike (dash-dotted)
 spin pair productions is also shown.}
\label{fig_cross-inv_SU}
\end{center}
\end{figure}

\begin{figure}[ht]
\begin{center}
  \epsfxsize=12cm
  \epsfbox{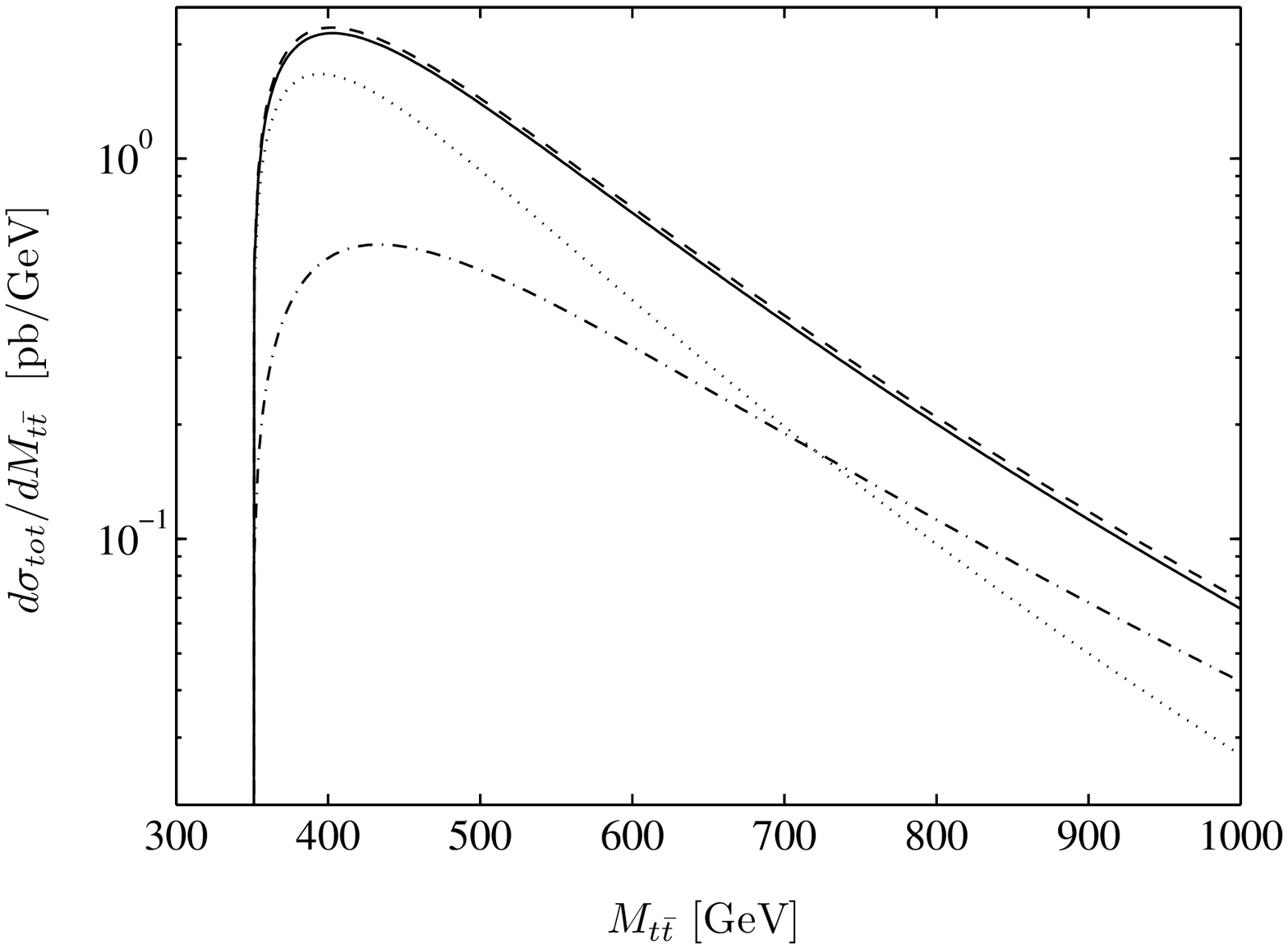}
\caption{Differential cross section \p{total_s} as a function 
 of the top-antitop invariant mass $M_{t\bar{t}}$ for the SM $+$ vector 
 unparticle processes 
 with $d_{\cal U}=1.20$ and $\Lambda=1$ TeV. 
The solid line depicts the result of the SM and the 
 dashed line depicts the result of the SM $+$ vector unparticle.
Breakdown of the latter into the like (dotted) and the unlike (dash-dotted)
 spin pair productions is also shown.}
\label{fig_cross-inv_VU}
\end{center}
\end{figure}

With the lower bound on $d_{\cal U}$ from the Tevatron experiment, 
 we now consider the unparticle effects on top-antitop production 
 process at the LHC. 
The dependence of the cross section 
 on the top-antitop invariant mass $M_{t\bar{t}}$ is given by 
\begin{eqnarray}
 \frac{d \sigma_{tot}(pp \rightarrow t\bar{t})}{d\sqrt{s}}=
  \sum_{a,b} \int\limits_{-1}^{1} d \cos\theta
   \int\limits_{\frac{s}{E_{CMS}^2}}^1 d x_1 
   \frac{2\sqrt{s}}{x_1 E_{CMS}^2} f_a(x_1,Q^2)
   f_b\left(\frac{s}{x_1 E_{CMS}^2},Q^2\right)
   {d\sigma(t\bar{t}) \over d\cos \theta}.
   \label{total_s}
\end{eqnarray}
Figs. \ref{fig_cross-inv_SU} and \ref{fig_cross-inv_VU} show the same dependence
 for the case of the scalar and the vector unparticles.
Here, the decomposition of the total cross section into the like 
 ($t_{\uparrow}\bar{t}_{\uparrow}+t_{\downarrow}\bar{t}_{\downarrow}$)
 and unlike ($t_{\uparrow}\bar{t}_{\downarrow}+t_{\downarrow}\bar{t}_{\uparrow}$)
 top-antitop spin pairs is also shown.

\begin{figure}[ht]
\begin{center}
  \epsfxsize=12cm
  \epsfbox{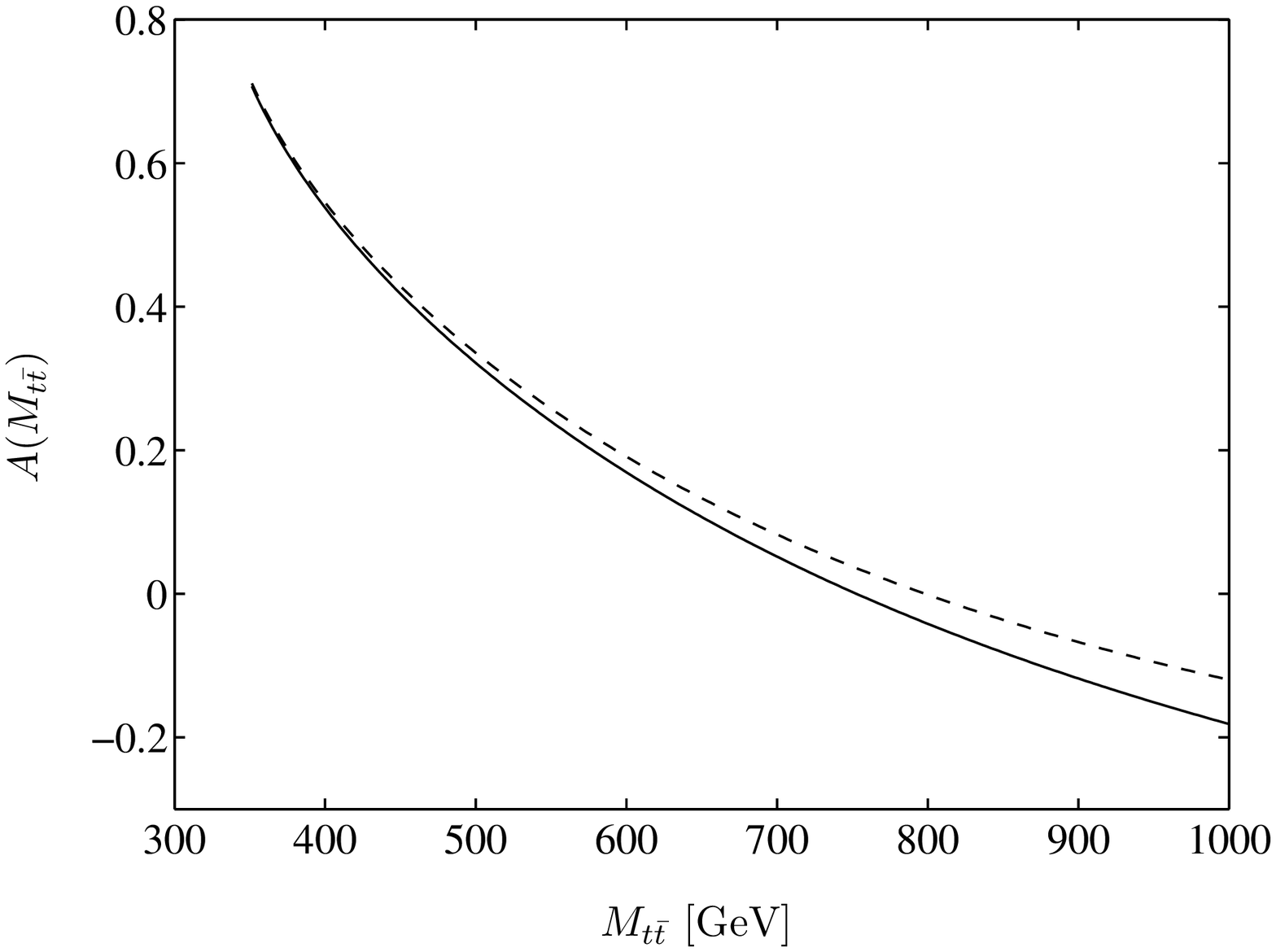} \\
\caption{Spin asymmetry ${\cal A}$ as a function of the top-antitop
 invariant mass $M_{t\bar{t}}$ 
 with $d_{\cal U}=1.01$ and $\Lambda=1$ TeV. 
The solid line corresponds to the SM,
 while the dashed line corresponds to the result of the SM $+$ scalar unparticle.}
\label{fig_A_s_SU}
\end{center}
\end{figure}

\begin{figure}[ht]
\begin{center}
  \epsfxsize=12cm
  \epsfbox{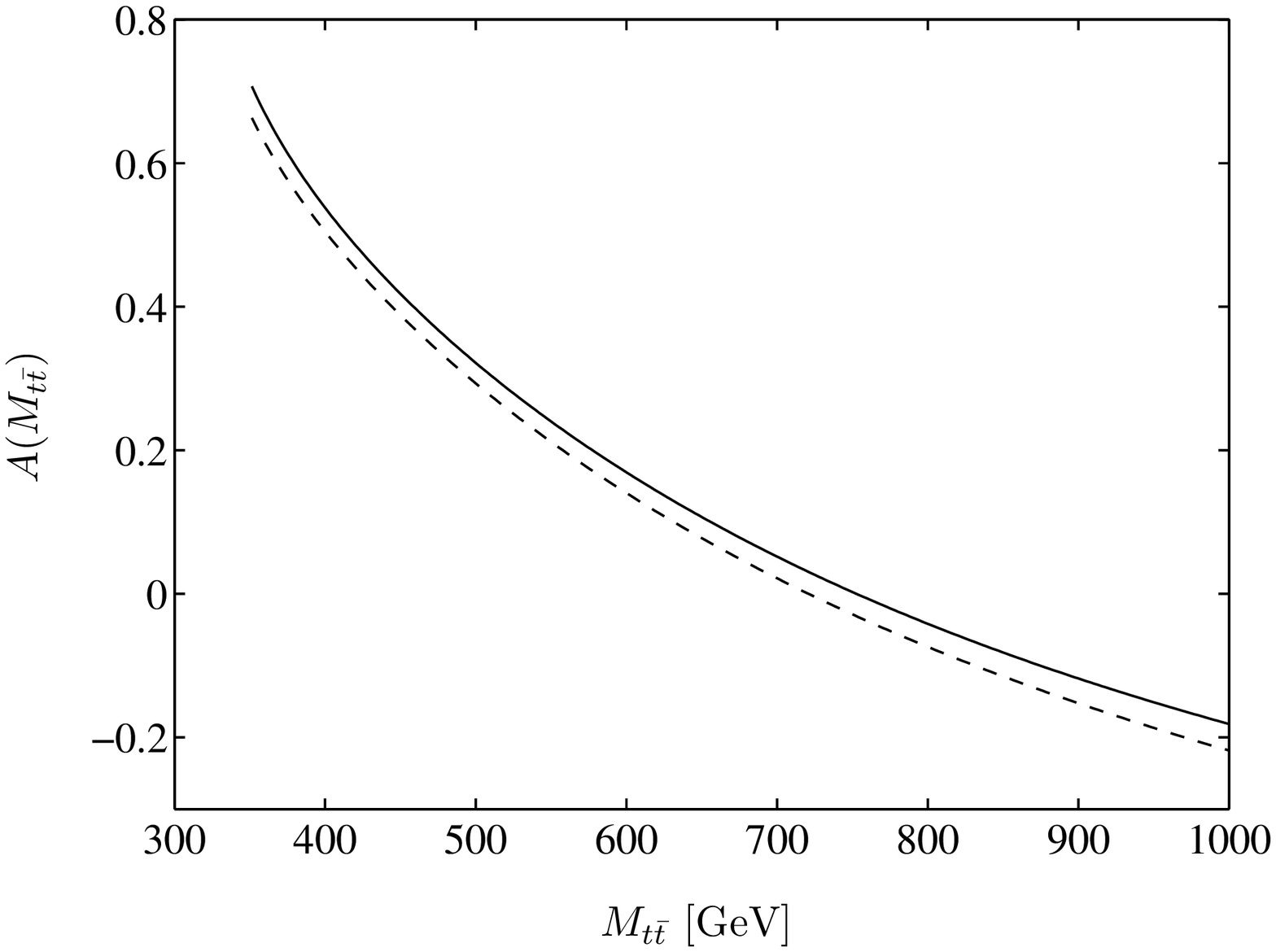} \\
\caption{Spin asymmetry ${\cal A}$ as a function of the top-antitop
 invariant mass $M_{t\bar{t}}$
 with $d_{\cal U}=1.20$ and $\Lambda=1$ TeV. 
The solid line corresponds to the SM,
 while the dashed line corresponds to the result of the SM $+$ vector unparticle.}
\label{fig_A_s_VU}
\end{center}
\end{figure}

\begin{figure}[ht]
\begin{center}
  \epsfxsize=12cm
  \epsfbox{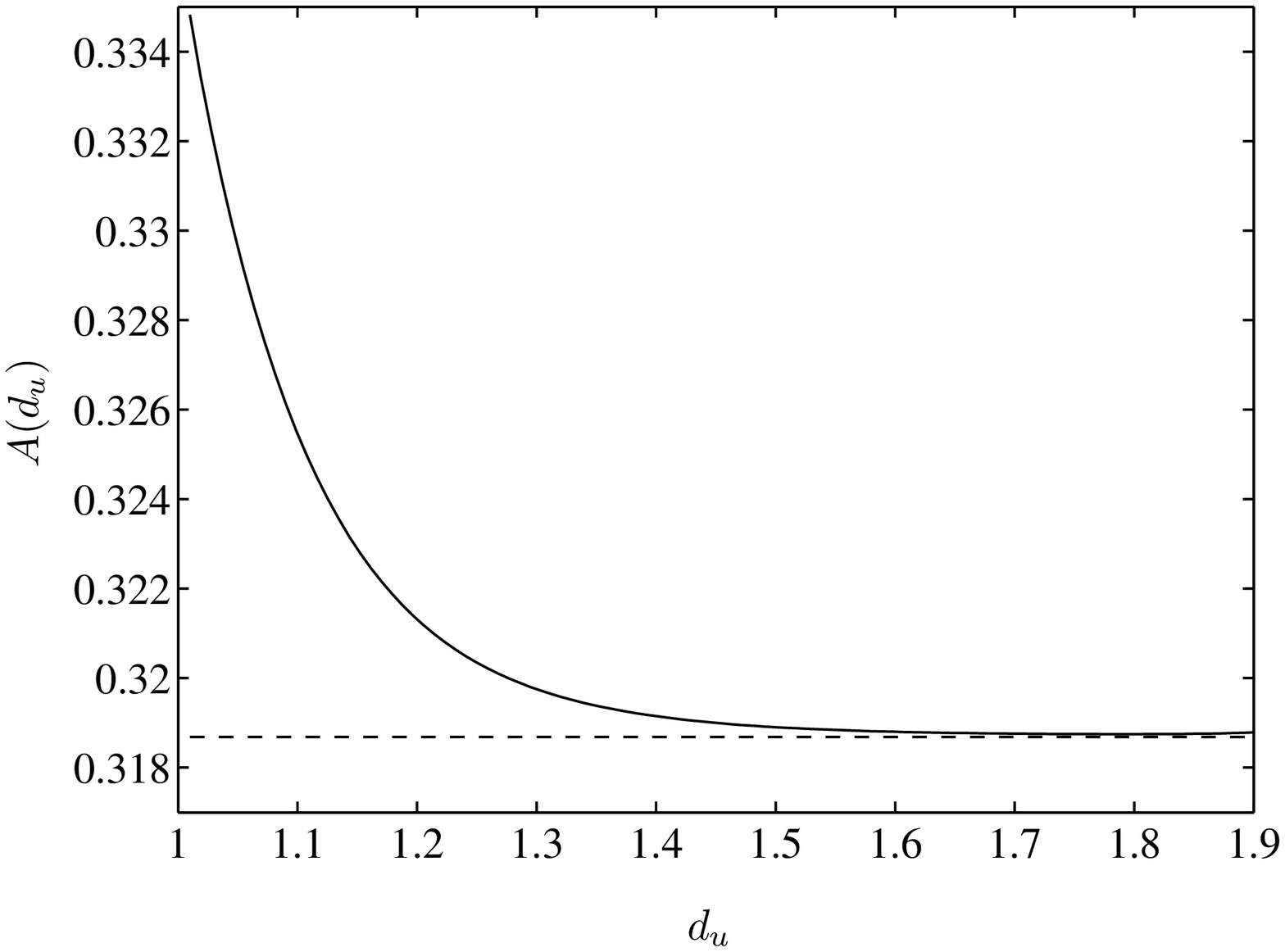} \\
\caption{Spin asymmetry ${\cal A}$ as a function of $d_{\cal U}$
 for the case of the scalar unparticle 
 with $\Lambda=1$ TeV.}
\label{fig_A_du_SU}
\end{center}
\end{figure}

\begin{figure}[ht]
\begin{center}
  \epsfxsize=12cm
  \epsfbox{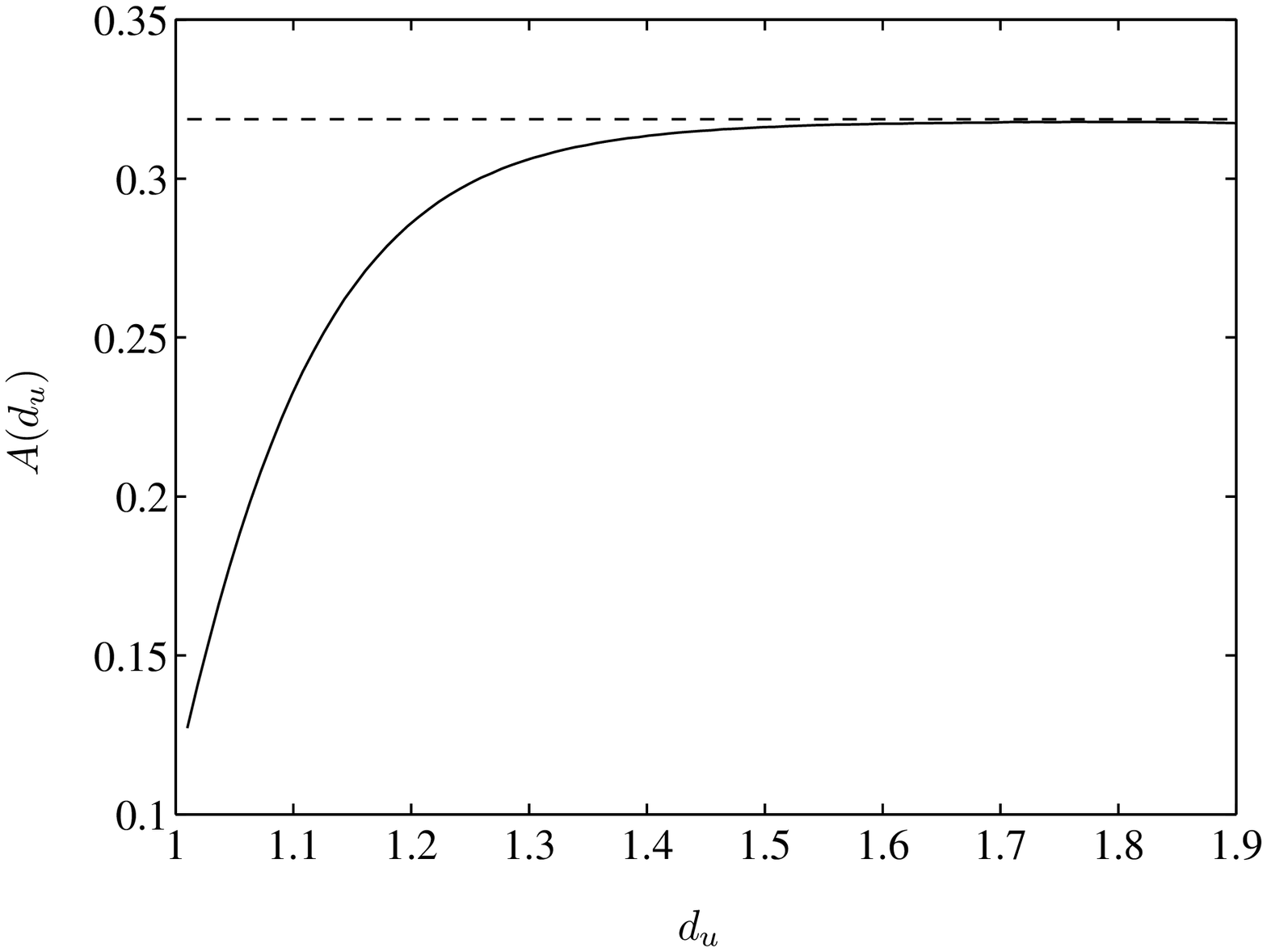} \\
\caption{Spin asymmetry ${\cal A}$ as a function of $d_{\cal U}$
 for the case of the vector unparticle 
 with $\Lambda=1$ TeV.
}
\label{fig_A_du_VU}
\end{center}
\end{figure}

\begin{figure}[ht]
\begin{center}
  \epsfxsize=12cm
  \epsfbox{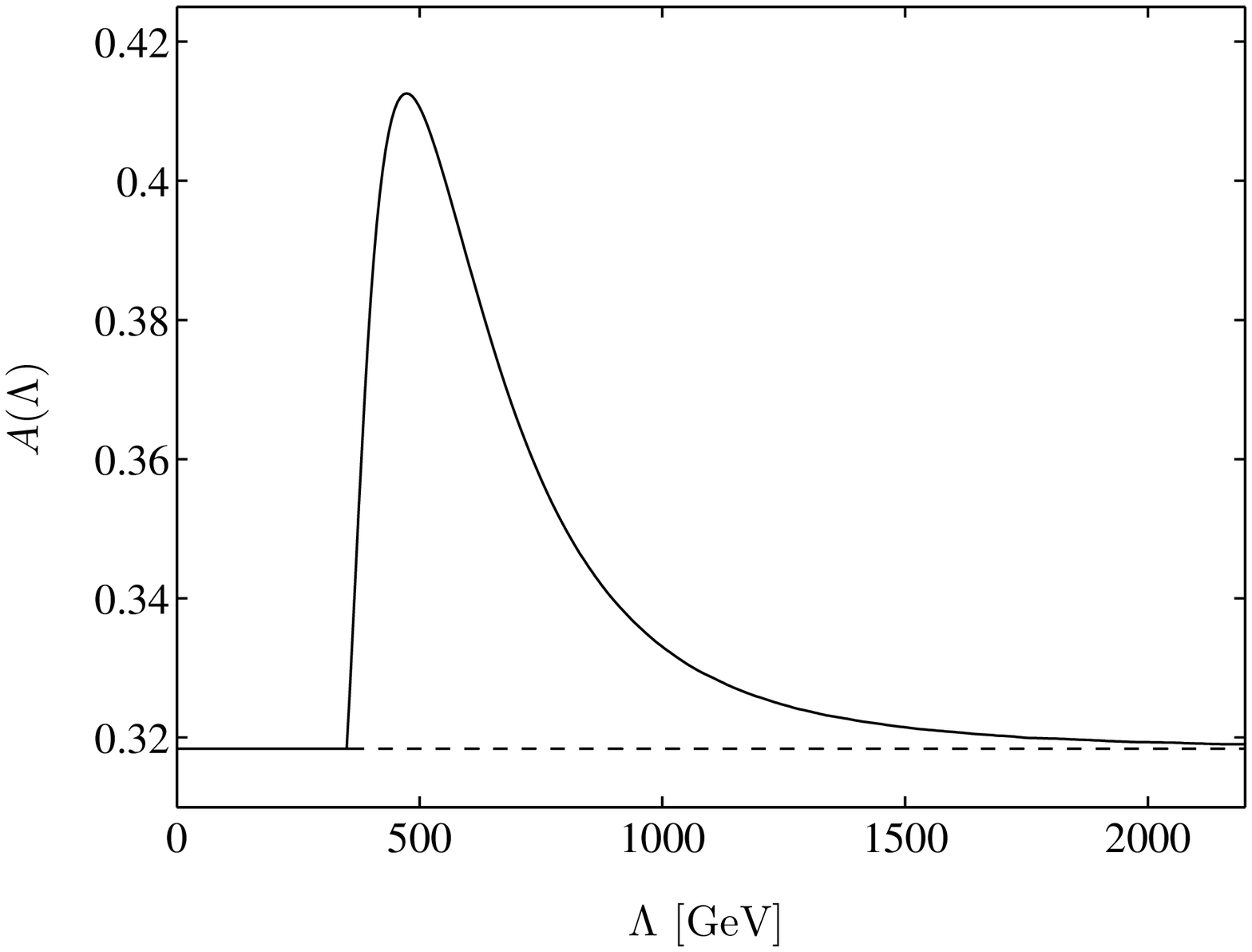} \\
\caption{Spin asymmetry ${\cal A}$ as a function of $\Lambda$
 for the case of the scalar unparticle 
 with $d_{\cal U}=1.01$.
}
\label{fig_A_Lambda_SU}
\end{center}
\end{figure}

\begin{figure}[ht]
\begin{center}
  \epsfxsize=12cm
  \epsfbox{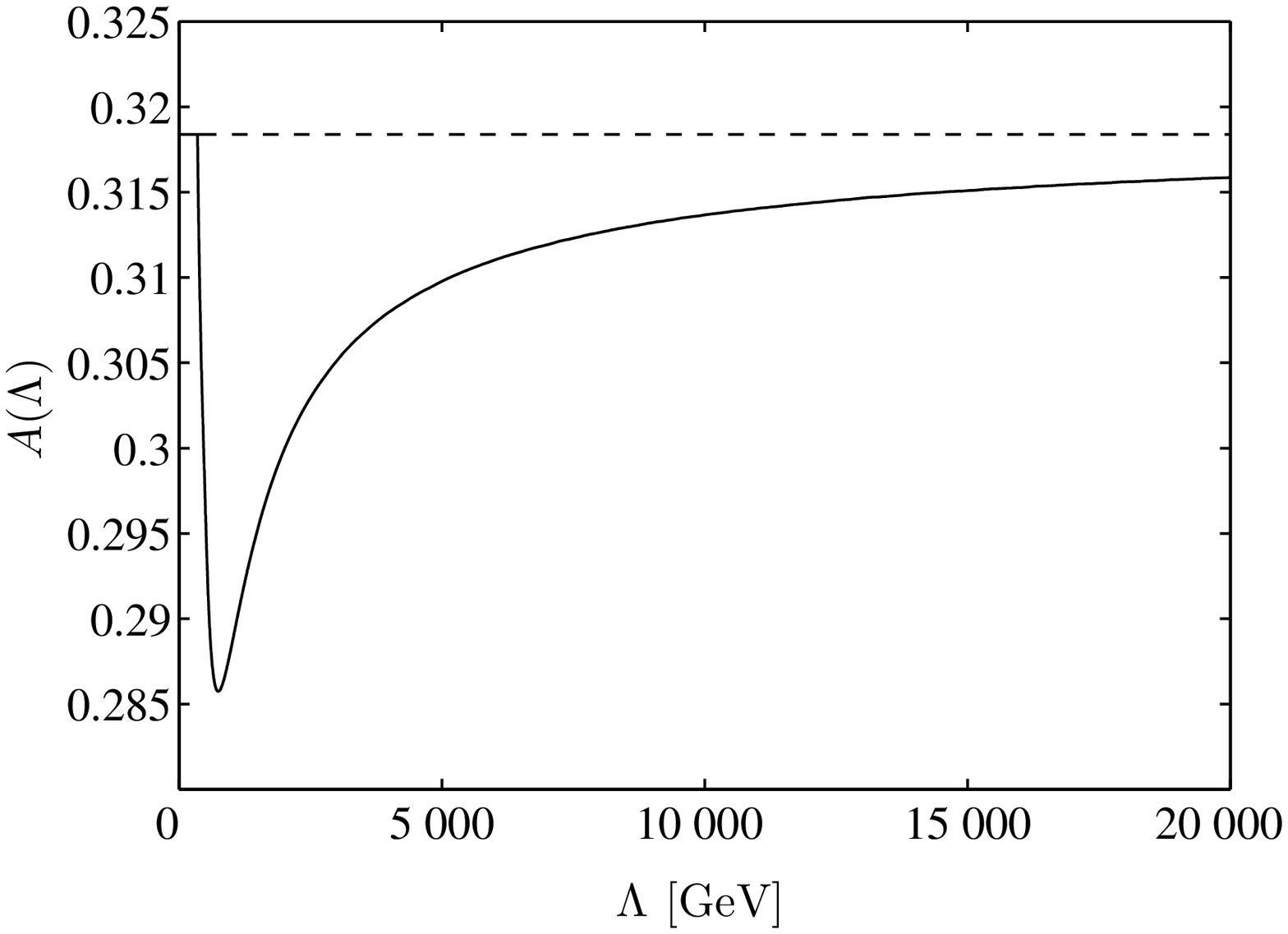} \\
\caption{Spin asymmetry ${\cal A}$ as a function of $\Lambda$
 for the case of the vector unparticle 
 with $d_{\cal U}=1.20$.
}
\label{fig_A_Lambda_VU}
\end{center}
\end{figure}

Now we show the results for the spin asymmetry $\cal A$ as a function of the top-antitop
 invariant mass $M_{t\bar{t}}$. 
The plot for the case of the scalar unparticle is shown in Fig. \ref{fig_A_s_SU}
 and for the vector unparticle in Fig. \ref{fig_A_s_VU}.
The dependence of $\cal A$ on the value of $d_{\cal U}$, after integration with respect
 to $M_{t\bar{t}}$ in the range $2m_t \leq M_{t\bar{t}} \leq E_{CMS}$,
 is depicted in Figs. \ref{fig_A_du_SU} and \ref{fig_A_du_VU}.
The existence of the scalar unparticle increases the value of $\cal A$, while the
 existence of the vector unparticle decreases expected spin asymmetry.
Deviation from the SM becomes larger as the center-of-mass energy becomes larger and
 $d_{\cal U}$ becomes smaller.
In Figs. \ref{fig_A_Lambda_SU} and \ref{fig_A_Lambda_VU}, 
 the results for the spin asymmetry $\cal A$ as a
 function of the effective cutoff scale $\Lambda$ are depicted.
Again one can see that the scalar(vector) unparticle gives rise 
 to positive(negative) contribution to ${\cal A}$.

Table \ref{table} presents values of the spin asymmetry $\cal A$ and $t\bar{t}$ total 
 cross section for selected values of $d_{\cal U}$. 
Here, ${\cal A}^{\rm (cut)}$ and $\sigma^{\rm (cut)}$ denote 
 the results when we take into account the unparticle contributions 
 only for the range $\sqrt{s}=M_{t\bar{t}}\le\Lambda$. 
For the spin asymmetry $\cal A$, we see deviation $\sim 5.0\%$ 
 for the allowed value of $d_{\cal U} = 1.01$ for the
 model with the scalar unparticle, and $\sim 10\%$ 
 for the allowed value of $d_{\cal U} = 1.20$ 
 for the model with the vector unparticle. 
Note that for a fixed $d_{\cal U}$, 
 the deviation of the spin asymmetry from the SM one 
 is always bigger than the deviation of the total cross sections. 
With the estimated precision of the measurement around 6$\%$ \cite{ATLAS_TOP}, 
 the size of the deviation from the SM for the vector unparticle 
 could be sufficient for the observation in the data from the first year 
 of low luminosity LHC run
 (with integral luminosity ${\cal L} = 10$ fb$^{-1}$). 
For higher values of $d_{\cal U}$, the interactions between the
 unparticles and the SM are suppressed, and thus the deviation 
 is very small. 
In particular, the vector unparticle with the rigid conformal
 invariance, where $d_{\cal U} \geq 3$, 
 does not give rise to large deviations.

\begin{table}[t]
\begin{center}
\begin{tabular}{c c c c c}
\hline
$d_{\cal U}$  & ${\cal A}_{\rm SU}$ & $\sigma_{\rm SU}$ [pb] 
 & ${\cal A}_{\rm SU}^{\rm (cut)}$ & $\sigma_{\rm SU}^{\rm (cut)}$ [pb] \\
\hline
 1.01 & 0.335 & 502 & 0.333 & 501   \\
 1.10 & 0.325 & 495 & 0.324 & 494   \\
\hline
$d_{\cal U}$ 
 & ${\cal A}_{\rm VU}$ & $\sigma_{\rm VU}$ [pb]
 & ${\cal A}_{\rm VU}^{\rm (cut)}$ & $\sigma_{\rm VU}^{\rm (cut)}$ [pb] \\
\hline
 1.20 & 0.286 & 508 & 0.288 & 506 \\
 3.01 & 0.318 & 490 & 0.318 & 490 \\
\hline
 & ${\cal A}$ & $\sigma$ [pb] &  & \\
\hline
 SM & 0.319 & 489 &  & \\
\hline
\end{tabular}
\end{center}
\caption{
Spin asymmetry ${\cal A}$ and $t\bar{t}$ total cross section
 for the top-antitop events without the constraint on the invariant mass
 (second and third column) 
 and with the invariant mass cut (fourth and fifth column) in the range 
 $M_{t\bar{t}}\le\Lambda$ GeV.
The last line shows the SM results.
 }
\label{table}
\end{table}

\section{Conclusion}
We have studied the top-antitop pair production 
 and the top spin correlations with the scalar and the vector unparticles
 at the LHC.
In addition to the SM processes, 
 there is a new contribution to the top-antitop pair production 
 process mediated by unparticles in the $s$-channel.
We have computed the corresponding density matrix 
 for the top-antitop pair production including the contributions
 by the scalar unparticle and the vector unparticle exchanges. 
The scalar unparticle contributes to the like spin pair production 
 amplitude through the gluon fusion processes, while the vector 
 unparticle mainly contributes to the unlike spin pair 
 through the quark annihilation process.

We have shown various numerical results for 
 the production cross sections and the top spin correlations 
 with certain values of the scaling dimension $d_{\cal U}$
 and the cutoff $\Lambda$.
We have found a sizable deviation of the top-antitop pair production 
 cross sections and the top spin correlations 
 from those in the SM for the scalar and vector unparticle exchange 
 processes with lower values of the
 scaling dimensions $d_{\cal U}$.
In particular, for the spin correlation, we have found about $5.0\%$
 and $10\%$ deviations of the spin asymmetry from the SM one 
 for the scalar unparticle and the vector unparticle, respectively.
In Ref. \cite{ATLAS_TOP}, it is shown that the spin asymmetry of 
 the top-antitop pairs in the SM will be measured with a precision of 
 $6\%$ after one LHC year at low luminosity, $10$ fb${}^{-1}$. 
Thus, the deviation of the top spin symmetry by the vector unparticle 
 effects can be measurable. 
However, note that it is very rough estimation 
 since the sensitivity of the ATLAS experiment on
 the spin correlation published in \cite{ATLAS_TOP} was estimated 
 selecting low energetic top quarks with $M_{t\bar{t}} < 550$ GeV.
In order to estimate the sensitivity more accurately with a high
 $M_{t\bar{t}}$ region for our case, we need elaborate Monte-Carlo 
 simulations including the detector response.
We leave this interesting subject for future study. 
\\\\
%
%
\noindent{\Large \bf Acknowledgements}\\
Authors would like to thank Santosh Kumar Rai  
 and Vladislav \v{S}im\'ak for illuminating discussion.
The work of M.A. is supported by the Science Research Center Program of
 the Korea Science and Engineering Foundation through the Center for
 Quantum Spacetime (CQUeST) of Sogang University with grant number
 R11-2005-021.
M.A. would also like to thank to Czech Technical University in Prague
 and theory division, KEK, for their 
 hospitality during his visit. 
The work of N.O. is supported in part by the Grant-in-Aid 
 for Scientific Research from the Ministry of Education, 
 Science and Culture of Japan (No. 18740170).
The work of K.S. is supported by the Research Program  
 MSM6840770029 and by the project of International Cooperation ATLAS-CERN 
 of the Ministry of Education, 
 Youth and Sports of the Czech Republic.
\newpage
%
%
\noindent{\Large \bf Appendix}
\appendix
\section{Couplings and decay widths}\label{appendix-A}
The couplings for the SM $Z$ boson:
\begin{eqnarray}
 g_{L,1}^\nu&=&{e \over \cos\theta_W\sin\theta_W}{1 \over
  2}\,,~~~g_{R,1}^\nu=0\,,\\
 g_{L,1}^l&=&{e \over \cos\theta_W\sin\theta_W}
   \left(-{1 \over 2}-\sin^2\theta_W(-1)\right)\,,
  ~~~g_{R,1}^l=-e(-1)\tan\theta_W\,,\\
 g_{L,1}^u&=&{e \over \cos\theta_W\sin\theta_W}
   \left({1 \over 2}-\sin^2\theta_W{2 \over 3}\right)\,,
  ~~~g_{R,1}^u=-e{2 \over 3}\tan\theta_W\,,\\
 g_{L,1}^d&=&{e \over \cos\theta_W\sin\theta_W}
   \left(-{1 \over 2}-\sin^2\theta_W\left(-{1 \over 3}\right)\right)\,,
  ~~~g_{R,1}^d=-e\left(-{1 \over 3}\right)\tan\theta_W\,.
\end{eqnarray}
The decay width of $Z$ boson:
\begin{eqnarray}
 \Gamma(Z\rightarrow \nu\bar{\nu})&=&{M_{Z} \over
  24\pi}((g_{L}^{\nu})^2+(g_{R}^{\nu})^2)\,,\\
 \Gamma(Z\rightarrow l\bar{l})&=&{M_{Z} \over
  24\pi}((g_{L}^{l})^2+(g_{R}^l)^2)\,,\\
 \Gamma(Z\rightarrow u\bar{u})&=&{M_{Z} \over
  24\pi}3((g_{L}^{u})^2+(g_{R}^u)^2)\,,\\
 \Gamma(Z\rightarrow d\bar{d})&=&{M_{Z} \over
  24\pi}3((g_{L}^d)^2+(g_{R}^d)^2)\,.
\end{eqnarray}
%
%
%

\newpage
%
%






\begin{thebibliography}{100}
%
\bibitem{Georgi1}
  H.~Georgi,
  Phys.\ Rev.\ Lett.\  {\bf 98} (2007) 221601
  [arXiv:hep-ph/0703260].
%
\bibitem{Georgi2}
  H.~Georgi,
  Phys.\ Lett.\  B {\bf 650} (2007) 275
  [arXiv:0704.2457 [hep-ph]].
%
\bibitem{Banks}
  T.~Banks and A.~Zaks,
  Nucl.\ Phys.\  B {\bf 196} (1982) 189.
%
\bibitem{ChKeYu1}
  K.~Cheung, W.~Y.~Keung and T.~C.~Yuan,
  Phys.\ Rev.\ Lett.\  {\bf 99} (2007) 051803
  [arXiv:0704.2588 [hep-ph]].
%
\bibitem{Abe}
  F.~Abe {\it et al.}  [CDF Collaboration],
  Phys.\ Rev.\ Lett.\  {\bf 74} (1995) 2626
  [arXiv:hep-ex/9503002].
%
\bibitem{Bigi}
I.~I.~Y.~Bigi, Y.~L.~Dokshitzer, V.~A.~Khoze, J.~H.~Kuhn and P.~M.~Zerwas,
  Phys.\ Lett.\  B {\bf 181} (1986) 157.
%
\bibitem{Stelzer}
  T.~Stelzer and S.~Willenbrock,
  Phys.\ Lett.\ B {\bf 374} (1996) 169 
  [arXiv:hep-ph/9512292];
  A.~Brandenburg,
  Phys.\ Lett.\ B {\bf 388} (1996) 626 
  [arXiv:hep-ph/9603333];
  D.~Chang, S.~C.~Lee and A.~Sumarokov,
  Phys.\ Rev.\ Lett.\  {\bf 77} (1996) 1218 
  [arXiv:hep-ph/9512417].
\bibitem{Mahlon-Parke}
  G.~Mahlon and S.~J.~Parke,
  Phys.\ Rev.\ D {\bf 53} (1996) 4886 
  [arXiv:hep-ph/9512264];
  Phys.\ Lett.\ B {\bf 411} (1997) 173 
  [arXiv:hep-ph/9706304].
%
\bibitem{Bernreuther2}
  W.~Bernreuther, A.~Brandenburg, Z.~G.~Si and P.~Uwer,
  Phys.\ Rev.\ Lett.\  {\bf 87} (2001) 242002 
  [arXiv:hep-ph/0107086];
  Nucl.\ Phys.\ B {\bf 690} (2004) 81 
  [arXiv:hep-ph/0403035].
%
\bibitem{ee}
  K.~Y.~Lee, H.~S.~Song, J.~H.~Song and C.~Yu,
  Phys.\ Rev.\  D {\bf 60} (1999) 093002
  [arXiv:hep-ph/9905227];
%
  K.~Y.~Lee, S.~C.~Park, H.~S.~Song and C.~Yu,
  Phys.\ Rev.\  D {\bf 63} (2001) 094010
  [arXiv:hep-ph/0011173];
%
  C.~X.~Yue, L.~Wang, L.~N.~Wang and Y.~M.~Zhang,
  Chin.\ Phys.\ Lett.\  {\bf 23} (2006) 2379.
%
\bibitem{gamma}
 K.~Y.~Lee, S.~C.~Park, H.~S.~Song, J.~H.~Song and C.~Yu,
  Phys.\ Rev.\  D {\bf 61} (2000) 074005
  [arXiv:hep-ph/9910466].
%
\bibitem{AOSS1}
  M.~Arai, N.~Okada, K.~Smolek and V.~\v{S}im\'ak,
  Phys.\ Rev.\ D {\bf 70} (2004) 115015
  [arXiv:hep-ph/0409273].
%
\bibitem{AOSS2}
  M.~Arai, N.~Okada, K.~Smolek and V.~\v{S}im\'ak,
  Phys.\ Rev.\  D {\bf 75} (2007) 095008
  [arXiv:hep-ph/0701155].
%
\bibitem{AOSS3}
  M.~Arai, N.~Okada, K.~Smolek and V.~\v{S}im\'ak,
  Acta. Phys. Polon. B {\bf 40} (2009) 93
  [arXiv:0804.3740 [hep-ph]].
%
\bibitem{maltoni}
  R.~Frederix and F.~Maltoni,
  JHEP {\bf 0901} (2009) 047
  [arXiv:0712.2355 [hep-ph]].
%
\bibitem{ChGh}
  D.~Choudhury and D.~K.~Ghosh,
  Int.\ J.\ Mod.\ Phys.\  A {\bf 23} (2008) 2579
  [arXiv:0707.2074 [hep-ph]].
%
\bibitem{AlPa}
  A.~T.~Alan and N.~K.~Pak,
  Europhys.\ Lett.\  {\bf 84} (2008) 11001
  [arXiv:0708.3802 [hep-ph]].
%
\bibitem{LiLiSiYa}
  H.~F.~Li, H.~L.~Li, Z.~G.~Si and Z.~J.~Yang,
  ``Unparticle Effects on Top Quark Pair Production at Photon Collider,''
  arXiv:0802.0236 [hep-ph].
%
\bibitem{Sa1}
  B.~Sahin,
  ``Unparticle Effects on Top Quark Spin Correlations in $e^+e^-$ Collision,''
  arXiv:0802.1937 [hep-ph].
%
\bibitem{Sa2}
  I.~Sahin,
  ``Effect of top quark spin on the unparticle couplings in $\gamma\gamma \to
  t\bar{t}$,''
  arXiv:0802.2818 [hep-ph].
%
\bibitem{Jezabek}
  A.~Czarnecki, M.~Jezabek and J.~H.~K\"uhn,
  Nucl.\ Phys.\ B {\bf 351} (1991) 70.
%
\bibitem{Bernreuther}
  W.~Bernreuther, O.~Nachtmann, P.~Overmann and T.~Schr\"oder,
  Nucl.\ Phys.\ B {\bf 388} (1992) 53, 
  [Erratum-ibid.\ B {\bf 406}, 516 (1993)];
  A.~Brandenburg and J.~P.~Ma,
  Phys.\ Lett.\ B {\bf 298} (1993) 211.
%
\bibitem{uwer}
  P.~Uwer,
  Phys.\ Lett.\  B {\bf 609} (2005) 271
  [arXiv:hep-ph/0412097].
%
\bibitem{CTEQ}
  J.~Pumplin, D.~R.~Stump, J.~Huston, H.~L.~Lai, P.~Nadolsky and W.~K.~Tung,
  JHEP {\bf 07} (2002) 012
  [arXiv:hep-ph/0201195].
%
\bibitem{Mack}
  G.~Mack and K.~Symanzik,
  Commun.\ Math.\ Phys.\  {\bf 27} (1972) 247.
%
\bibitem{Grinstein}
  B.~Grinstein, K.~Intriligator and I.~Z.~Rothstein,
  Phys.\ Lett.\  B {\bf 662} (2008) 367
  [arXiv:0801.1140 [hep-ph]].
%
\bibitem{ATLAS_TOP}
  F.~Hubaut, E.~Monnier, P.~Pralavorio, V.~\v{S}im\'ak, K.~Smolek,
  Eur.\ Phys.\ J.\ C {\bf 44} (2005) 13
  [arXiv:hep-ex/0508061].
%
\bibitem{CDF_cross}
  The CDF Collaboration,
  CDF note 9399.
%
\bibitem{Cacciari}
  M. Cacciari et al.,
  JHEP {\bf 404} (2004) 608 .
\end{thebibliography}
\end{document}